\documentclass[12pt, a4paper]{article}
\pdfoutput=1

\usepackage{amsmath}
\usepackage{amsfonts}
\usepackage{amssymb}
\usepackage{graphicx, rotating}
\usepackage{epsfig}
\usepackage{latexsym}
\usepackage{graphicx}
\usepackage{color}
\usepackage{amsmath,bm,amssymb}
\usepackage{cite}
\usepackage{slashed}
\usepackage{hyperref}
\hypersetup{colorlinks, citecolor=bluscuro, linkcolor=black, urlcolor=bluscuro}
\definecolor{rossos}{cmyk}{0,1,1,0.55}
\definecolor{bluscuro}{rgb}{0.15, 0.2, .85}
\definecolor{bluchiaro}{cmyk}{1,.3,0.,0.1}

\newcommand{\be}{\begin{equation}}
\newcommand{\ee}{\end{equation}}
\newcommand{\bea}{\begin{eqnarray}}
\newcommand{\eea}{\end{eqnarray}}

\newcommand{\arXiv}[2]{\href{http://arxiv.org/pdf/#1}{{\tt [#2/#1]}}}
\newcommand{\arXivold}[1]{\href{http://arxiv.org/pdf/#1}{{\tt [#1]}}}

\def\bma#1{\mbox{\boldmath{$#1$}}}

\begin{document}
\allowdisplaybreaks
\begin{titlepage}
\begin{flushright}
IFT-UAM/CSIC-19-114
\end{flushright}
\vspace{.3in}

\vspace{1cm}
\begin{center}
{\Large\bf\color{black} 
Tunneling Without Bounce  
} \\
\vspace{1cm}{
{\large J.R.~Espinosa}
} \\[7mm]
{\it Instituto de F\'{\i}sica Te\'orica UAM/CSIC, \\ 
C/ Nicol\'as Cabrera 13-15, Campus de Cantoblanco, 28049, Madrid, Spain
}
\end{center}
\bigskip

\vspace{.4cm}

\begin{abstract}
The false vacua of some potentials do not decay via Euclidean bounces. 
This typically happens for tunneling actions with a flat direction (in field configuration space) that is lifted by a perturbation into a sloping valley, pushing the bounce off to infinity.
Using three different approaches we find a consistent picture for such decays. In the Euclidean approach the bottom of the action valley consists of a family of pseudo-bounces (field configurations with some key good properties of bounces except extremizing the action). The pseudo-bounce result is validated by minimizing a WKB action in Minkowski space along appropriate paths in configuration space. Finally, the simplest approach uses the tunneling action method proposed recently with a simple modification of boundary conditions.
\end{abstract}
\bigskip

\end{titlepage}

\section{Introduction\label{sec:introduction}}

Metastable (false) vacua appear often in models of particle physics,
from beyond the Standard Model quantum field theories to the landscape of string theory. Perhaps even our Standard Model vacuum is metastable \cite{SM,Matthew,CMS}. Such vacua are also quite relevant in cosmology as our Universe might have spent time in them between different cosmological phase transitions.

Such false vacua can decay via quantum tunneling (by the nucleation of bubbles of a more stable phase that expand and transform the false vacuum into a deeper one). 
When the false vacuum is sufficiently long-lived, its decay rate 
(per unit volume) is exponentially suppressed and given by
\be
\Gamma/V = A\  e^{-S/\hbar} [1+{\cal O}(\hbar)]\ ,
\ee
where the exponential prefactor $A$ has dimensions of $[$energy$]^4 $ and the crucial quantity is the tunneling action $S$. We assume $S/\hbar\gg 1$ so that the semiclassical approximation applies.

\begin{figure}[t!]
\includegraphics[width=0.5\textwidth]{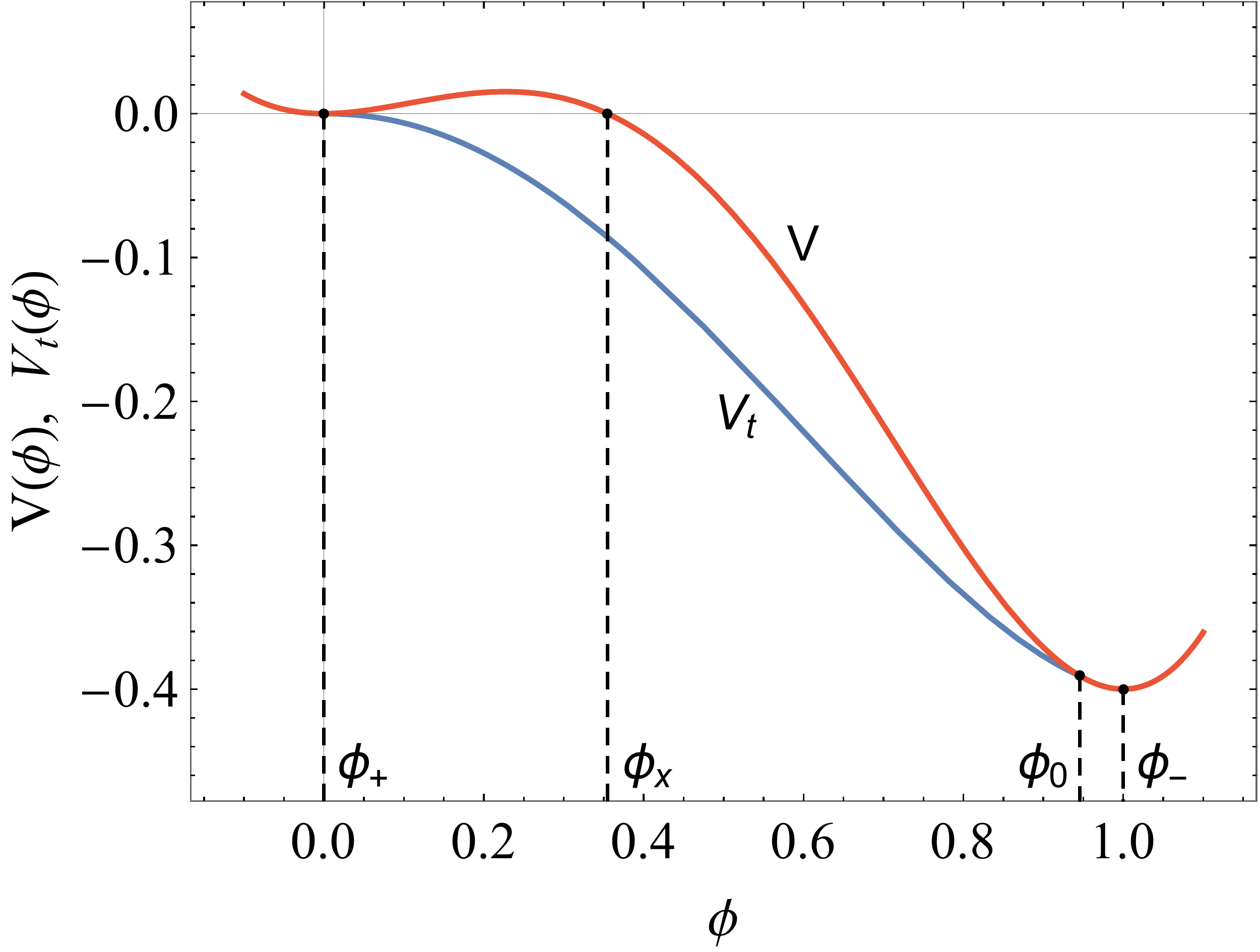}
\includegraphics[width=0.5\textwidth]{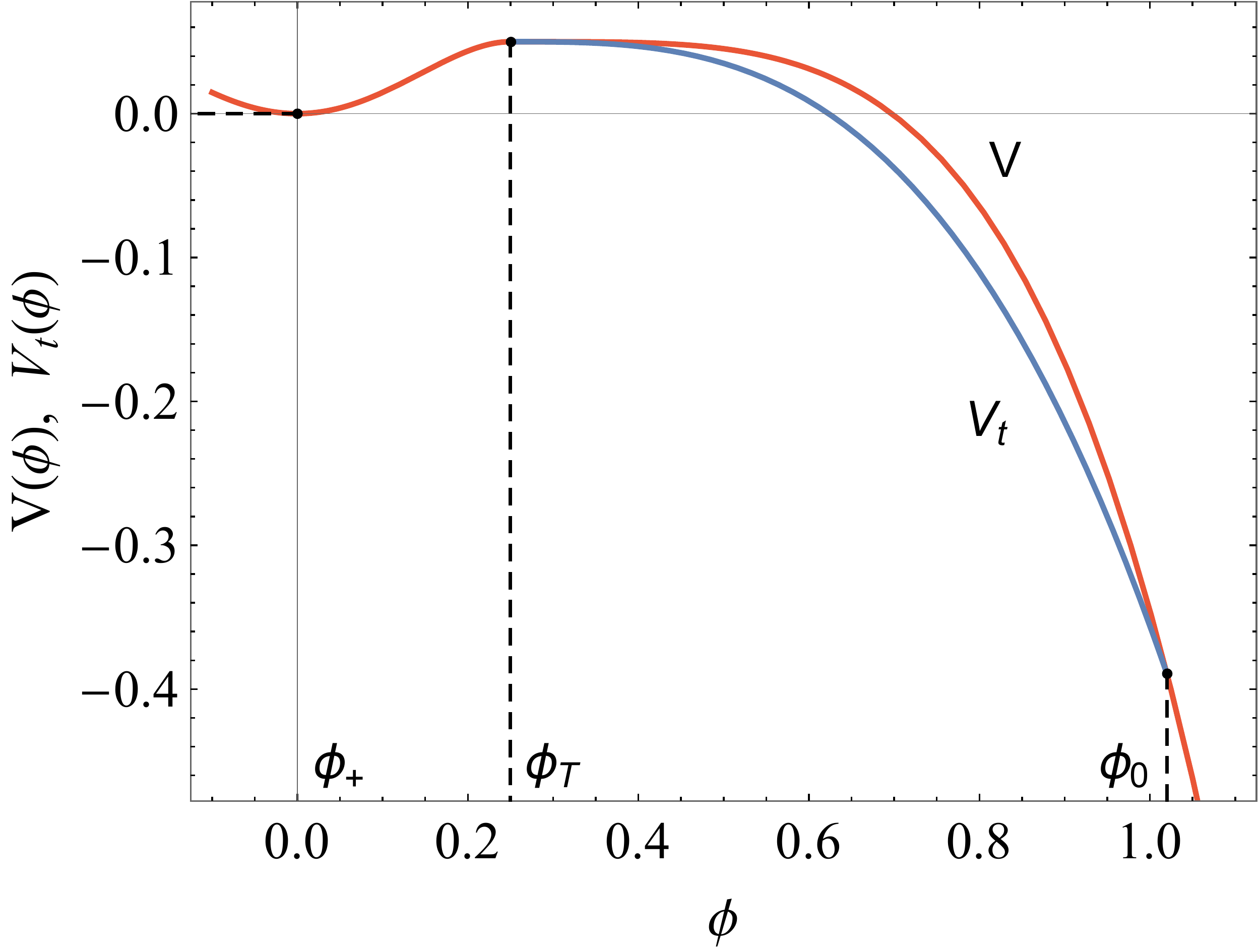}
\caption{Potentials $V(\phi)$ with a false vacuum at $\phi_+$ and its tunneling potentials, $V_t(\phi)$.
Left: Generic case. Right: Case without bounce.
\label{fig:Potential}
}
\end{figure}

There is a well known and elegant procedure, due to Coleman \cite{Coleman}, to calculate $S$. It goes as follows. 
Take for simplicity a single real scalar field $\phi$ in 4 dimensions, with a potential $V(\phi)$ that features a metastable local minimum at $\phi_+$ and a deeper minimum at $\phi_-$, see Fig.~\ref{fig:Potential}, left plot. Without loss of generality (and in the absence of gravity) fix $\phi_+=0$ and $V(\phi_+)=0$. 

The tunneling action $S$ for the decay of the false vacuum 
at $\phi_+$ is calculated by finding an $O(4)$-symmetric bounce 
$\phi_B(r)$ (or Euclidean bounce) that interpolates between the false vacuum $\phi_+$ and (the basin of) the true vacuum at $\phi_-$ and back to $\phi_+$. Such bounce extremizes \cite{CGM} the Euclidean action for the scalar field, which for $O(4)$-symmetric configurations reads
\be
S_E[\phi] = 2\pi^2\int_0^\infty  \left[\frac12 \dot{\phi}^2 + V(\phi)-V(\phi_+)\right]r^3dr\ .
\label{SE}
\ee
Therefore, the bounce is a solution of the corresponding Euler-Lagrange equation:
\be
\ddot{\phi} +\frac{3}{r}\dot{\phi} = V'\ ,
\label{EoM4}
\ee
where a dot (prime) represents a derivative with respect to $r$ ($\phi$). The bounce boundary conditions are
\be
\dot\phi_B(0)=0\ ,\quad \phi_B(\infty)=\phi_+\ .
\ee
If we identify $r$ with time, Eq.~(\ref{EoM4}) describes the classical motion of a point particle in the inverted potential $-V(\phi)$ subject to a velocity-dependent and time-decreasing friction force. The bounce solution can be found by  changing the value of the field at the center of the bounce, $\phi_B(r=0)\equiv\phi_0$, until the boundary condition at $r\rightarrow \infty$ is satisfied. The tunneling action for the decay of the $\phi_+$ vacuum is then given as $S=S_E[\phi_B]$.

If the potential does have a true vacuum then it is generally guaranteed, by  the so-called undershooting and overshooting method, that the bounce solution exists.  If one solves the differential equation (\ref{EoM4}) starting with a $\phi_0$ lower than $\phi_{\mathrm x}$ [for which $V(\phi_{\mathrm x})=0$, see Fig.~\ref{fig:Potential} left plot], friction ensures that the 
solution does not reach $\phi_+$ (this is an undershot). On the other hand, starting with $\phi_0$ arbitrarily close to the minimum $\phi_-$, the field will spend much time rolling very slowly at the beginning, friction will become more and more irrelevant and (Euclidean) energy conservation will ensure that the field reaches 
$\phi_+$ with non zero velocity: an overshot. By continuity between such extreme cases there must exist a $\phi_0$ that 
lands the field at $\phi_+$ with zero velocity, corresponding precisely to the bounce solution, which can then be found by interval bisection. 

Nevertheless, for some potentials this is not the whole story.
One special, and well known, example is the simple unstable quartic potential $V=-\lambda\phi^4/4$ (of relevance for the study of the stability of the  Standard Model Higgs potential). This potential has the property that solutions of (\ref{EoM4}) with arbitrary starting point $\phi_0$ are all bounces, as they reach $\phi_+$ at $r\rightarrow \infty$. These are the so-called Fubini bounces \cite{Fubini,Lipatov} and can be obtained analytically as
\be
\phi_B(r)=\frac{\phi_0}{1+ r^2/R^2}\ ,
 \quad \mathrm{with}\quad
\frac{1}{R^2}=\frac18 \lambda \phi_0^2\ ,
\label{Fubini}
\ee
and lead to the tunneling action
\be
S_E[\phi_B]=\frac{8\pi^2}{3\lambda}\ .
\label{SE0}
\ee
The fact that the potential is scale-invariant explains why no
particular $\phi_0$ is singled-out for the bounce and
leads to a degenerate family of instantons with arbitrary $\phi_0$ (and size $R\sim 1/\phi_0$) and why 
the tunneling action does not depend on $\phi_0$ \cite{Arnold}.\footnote{In more detail, 
scale invariance implies that, if $\phi_B(r)$ is a bounce solution,
then the rescaled $a\,\phi_B(ar)$ (with $a>0$) is also a bounce solution. As seen explicitly from (\ref{Fubini}), the rescaling amounts to a rescaling of $\phi_0\rightarrow a\phi_0$ (or, equivalenty, a rescaling of the bounce radius $R\rightarrow R/a$).}
In other words, the action functional $S_E[\phi]$ has a flat direction
in field configuration space, consisting of the family of Fubini bounces, where it takes the value (\ref{SE0}).

There are other potentials with a false vacuum for which no Euclidean bounce describes its decay.  It is easy to come up with examples of  such potentials.
Consider as an example the class of potentials with a false vacuum 
at $\phi_+=0$, with some kind of barrier that reaches its maximum 
at $\phi_T>0$, beyond which the potential is simply
$V(\phi>\phi_T)=-\lambda(\phi-\phi_T)^4/4$ \cite{NoBounce}, see Fig.~\ref{fig:Potential}, right plot. Due to the special properties
of the quartic potential mentioned above, it is clear that
any solution of equation (\ref{EoM4}) starting at $r=0$ with some $\phi(0)=\phi_0>\phi_T$ and $\dot\phi(0)=0$ ends at $\phi(\infty)=\phi_T$, never reaching $\phi_+$: thus this class of potentials indeed has no Euclidean bounce.

Another simple example of potential without bounce is
\be
V(\phi)=\frac12 m^2\phi^2-\frac{\lambda}{4}\phi^4\ ,
\ee
for either sign of $m^2$. This is most clearly seen \cite{Affleck}
as follows. Assume there is a bounce $\phi_B(r)$ and consider the rescaled field profile $\phi_a(r)\equiv a \phi_B(ar)$. The Euclidean action for the rescaled field, after changing the integration variable, reads
\be
S_E[\phi_a]=2\pi^2\int_0^\infty \left[\frac12 \left(\frac{d\phi_B}{dr}\right)^2-\frac14 \lambda\phi_B^4\right]r^3dr +\frac{2\pi^2}{a^2}
\int_0^\infty \left[\frac12 m^2\phi_B^2\right]r^3dr \ .
\ee
As $\phi_B(r)$ is by assumption a bounce, it extremizes the 
Euclidean action and, therefore, one shoud have $dS_E[\phi_a]/da=0$ at $a=1$, which translates, for $m^2\neq 0$, into the condition
\be
\int_0^\infty \phi_B^2r^3dr=0\ ,
\ee 
which can only be satisfied for $\phi_B(r)\equiv 0$. This contradicts
the initial assumption about the existence of a non-trivial bounce.

Such potentials without bounce are the main focus of this paper.
There is nothing mysterious or subtle about them: Quantum fluctuations on the false vacuum still nucleate bubbles that probe the unstable part of the potential, with decay rates that depend on the shape of the bubble. Nevertheless, these potentials have caused some confusion in previous literature (see \cite{NoBounce} for a recent example) and we believe there is room for improvement over the methods developed to deal with them in the past, like using the so-called constrained instantons \cite{Affleck,FY} or valley equations \cite{OldValley1,OldValley2,OldValley3,NewValley}.  

This paper revisits this problem relying first on the standard Euclidean approach to find pseudo-bounce field profiles with 
finite action that can mediate vacuum decay. This generic type of profiles is illustrated with full analytical control in the simple scale-invariant potential $V=-\lambda\phi^4/4$ in section~\ref{sec:lambdaphi4}. 
Resorting to the Minkowskian WKB approach we show in section~\ref{sec:wkb} that  such field configurations can indeed mediate vacuum decay and that the decay rate is correctly given by their Euclidean action.

Alternatively, tunneling actions can be calculated without using  Euclidean bounces, as done in the formulation of \cite{E}. This new approach introduces a tunneling potential $V_t$ to describe the decay process and formulates the action calculation as a simple variational problem in field space. It is then natural to explore how this new approach deals with the class of potentials that admit no bounce.  Section~\ref{sec:Vt} shows how the
new approach can be directly applied without modification to this class of potentials to obtain what we call ``restricted'' $V_t$'s (with fixed end-point $\phi_0$) and the tunneling action calculated by the same expression used for generic potentials that admit a bounce.
The results of this new approach agree with those obtained for Euclidean pseudo-bounces. 
Moreover, these findings compare favorably with the results obtained using the constrained instanton approach or valley methods, as is shown in section~\ref{sec:civ}. In particular, pseudo-bounces
inherit some of the good properties of proper bounces, not shared
in general by previous approaches.

Sections~\ref{sec:m2} and \ref{sec:m2lambdaphi4} extend the previous analyses to the simple potentials $V=-m^2\phi^2/2$ (for which pseudo-bounces coexist with a proper bounce) and $V=m^2\phi^2/2-\lambda\phi^4/4$, respectively. We find that pseudo-bounces might be relevant even for vacua whose decay is dominated by a proper bounce. In section~\ref{sec:analyticVt} we show how to construct 
potentials for which an analytical treatment of the restricted tunneling potentials can be performed. Section~\ref{sec:conclusions} presents a summary and outlook.

\section{The potential  $\bma{V(\phi)=-\lambda\phi^4/4}$. Euclidean approach\label{sec:lambdaphi4}}

\begin{figure}[t!]
\begin{center}
\includegraphics[width=0.6\textwidth]{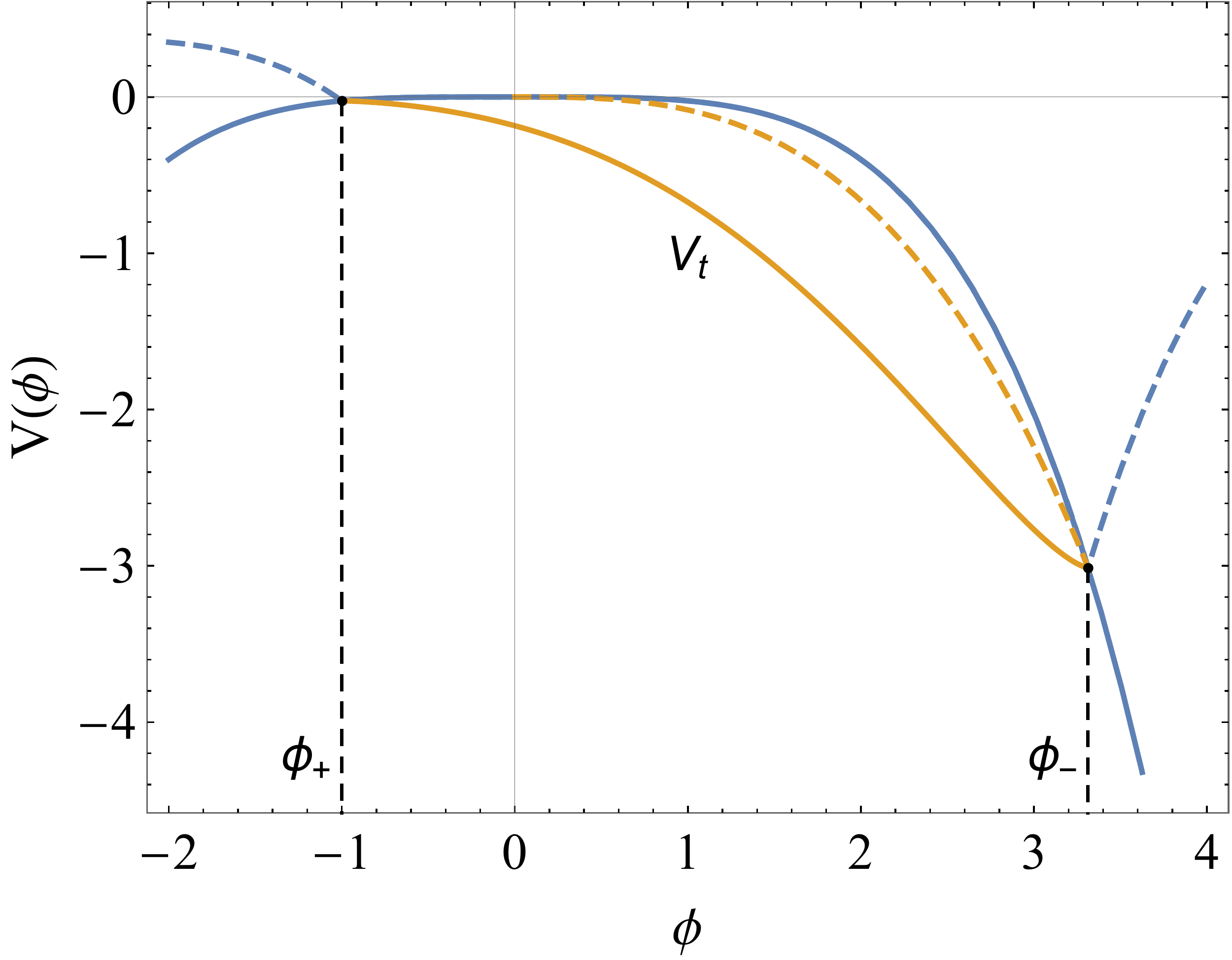}
\end{center}
\caption{Potential $V(\phi)=-\lambda\phi^4/4$ (blue line) for $\lambda=0.1$. Tunneling from $\phi_+$ to $\phi_-$
can be studied by modifying the potential as shown by the dashed lines (that create minima at $\phi_\pm$). The corresponding restricted tunneling potential, $V_t$,  for such decay is also plotted.
The dashed line corresponds to the standard $V_t$ that fails to reach $\phi_+$, as in Fig.~\ref{fig:Potential}. 
\label{fig:V}
}
\end{figure}

Instead of studying a no-bounce potential like the one 
in Fig.\ref{fig:Potential}, right plot, with an arbitrary barrier
from $\phi_+$ to $\phi_T$, it proves convenient to just take  $V=-\lambda\phi^4/4$. We modify this $V$ so that there is a false minimum at some $\phi_+<0$ and a deeper ``true'' vacuum at $\phi_->-\phi_+$ (without changing the potential between $\phi_+$ and $\phi_-$), see Fig.~\ref{fig:V}. We then consider the decay $\phi_+\rightarrow \phi_-$. 
 This gives the simplest no-bounce potential and the idea is to calculate $S_E(\phi_+\rightarrow \phi_-)$  and
see how this varies with growing $\phi_-$ for a fixed $\phi_+$.

Due to the simple form of the no-scale potential chosen we can readily guess some of the key properties of $S_E(\phi_+\rightarrow \phi_-)$. The only mass scales in the problem are $\phi_+$ and $\phi_-$ and therefore,
the dimensionless tunneling action (settting $\hbar=1$) must be a function of their ratio:
\be
S_E(\phi_+\rightarrow \phi_-)= f(-\phi_+/\phi_-)\ .
\ee
The ratio $-\phi_+/\phi_-\in (0,1)$ and the two boundary values of this interval are particularly simple.
For $\phi_-\rightarrow -\phi_+$ the two vacua become degenerate and the decay rate should vanish.
This implies
\be
f(1)=\infty\ .
\ee
The approach to this limiting value should be well described by the thin-wall approximation:
\be
S_{E,tw}=\frac{27\pi^2\sigma^4}{2\ \delta V^3}\ ,
\label{SEtw}
\ee
with the wall tension 
\be
\sigma= \int_{\phi_+}^{\phi_-}\sqrt{2[V(\phi)-V(\phi_-)]}\simeq  \frac{4}{3}\sqrt{\frac{\lambda}{2}}K(-1)\phi_-^3\ ,
\label{sigma}
\ee
where $K(-k^2)$ is the complete elliptic integral of the first kind
\be
K(-k^2)=\int_0^{\pi/2}\frac{1}{\sqrt{1+k^2\sin^2\theta}}d\theta ,
\label{Kk}
\ee 
and where
\be
\delta V\equiv V(\phi_+)-V(\phi_-)=\frac{\lambda}{4}(\phi_-^4-\phi_+^4)\ ,
\ee
is the energy difference between the vacua.\footnote{\label{twl}This thin-wall approximation can be refined further if the wall tension is 
defined as $\sigma= \int_{\phi_+}^{\phi_-}\sqrt{2[V(\phi)-V_l(\phi)]}$ where $V_l(\phi)$ is a constant slope potential connecting
$V(\phi_+)$ and $V(\phi_-)$.}

On the other hand, for $\phi_+=0$ and arbitrary $\phi_-$ we recover the case of Fubini bounces and therefore
\be
f(0)=\frac{8\pi^2}{3\lambda}\ .
\ee
The same limit should be reached asymptotically for $\phi_+\neq 0$ and $\phi_-\rightarrow \infty$. The absence of a bounce for $V$   indicates also that this limit will be reached from above, with $f$ decreasing monotonically with $-\phi_+/\phi_-\rightarrow 0$.

\begin{figure}[t!]
\includegraphics[width=0.5\textwidth]{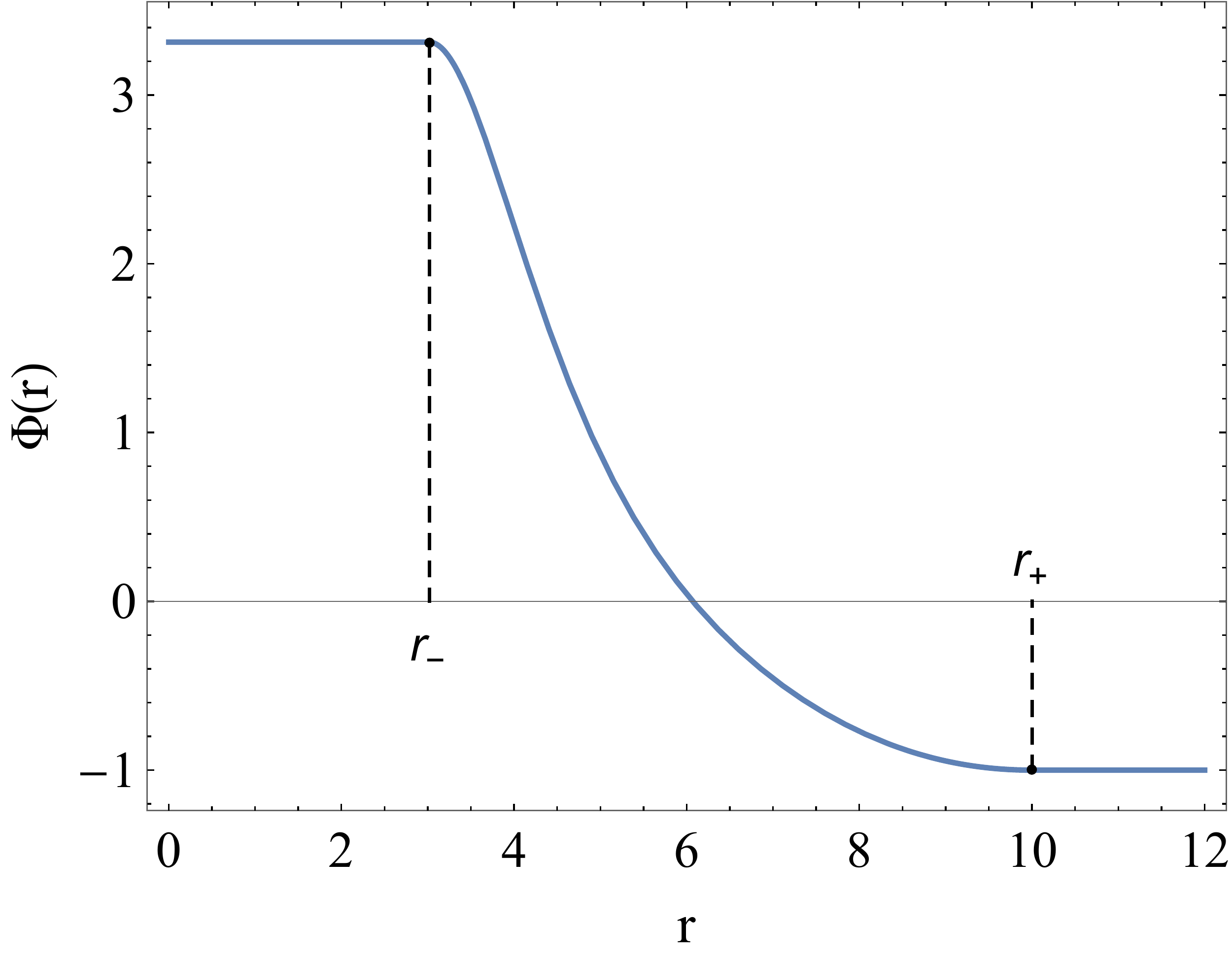}
\includegraphics[width=0.5\textwidth]{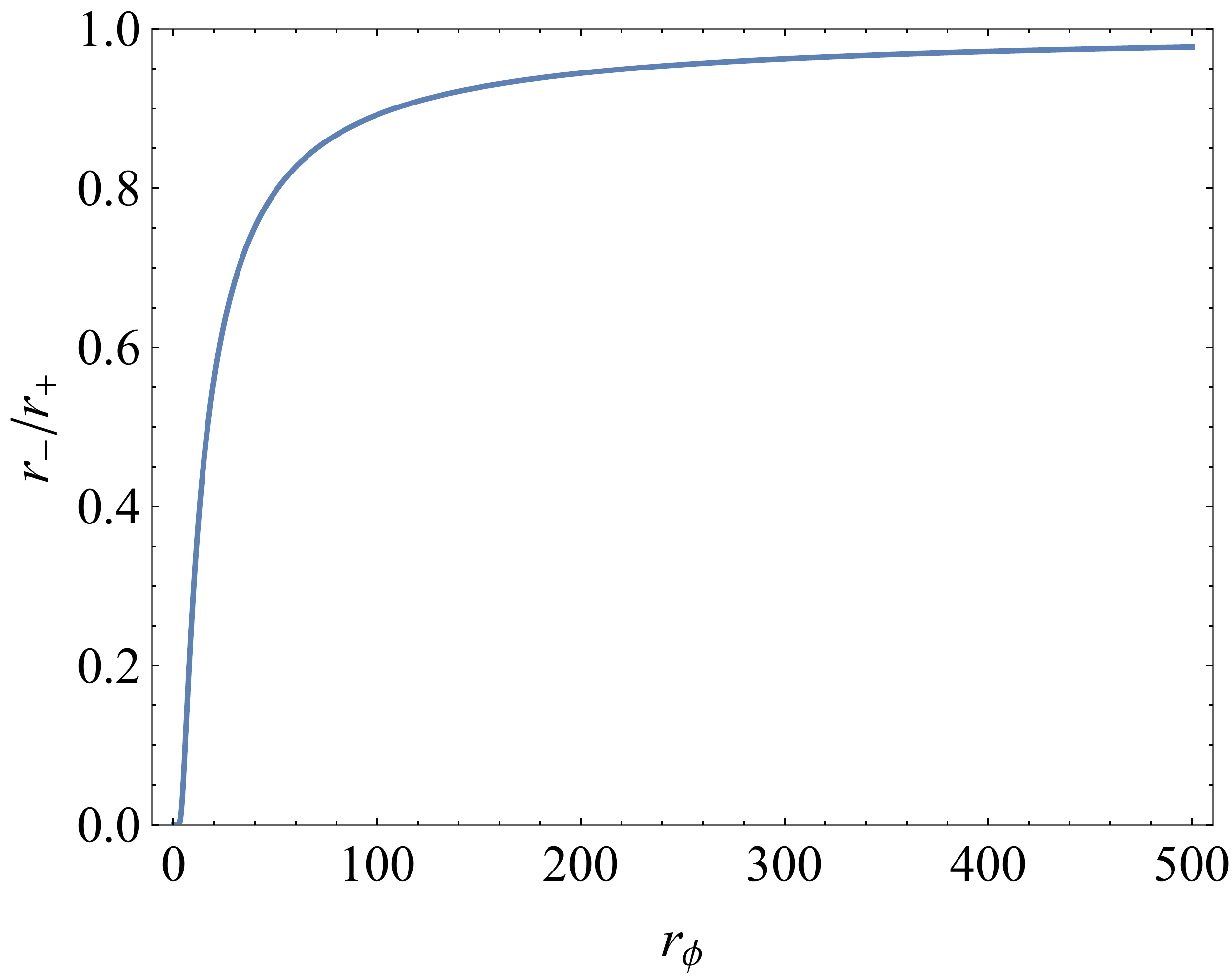}
\caption{Left: Profile of the Euclidean pseudo-bounce for the potential of Fig.~\ref{fig:V} 
as given by Eq.~(\ref{pseudobounce}), with $\phi_+=-1$ and $r_+=10$. Right: The ratio $r_-/r_+$ as a function of $r_\phi=-r_+\phi_+=r_-\phi_-$ as given by Eq.~(\ref{rphi}). For the case of the left plot, this gives $\phi_-=3.313$ and $r_-=3.02$.
\label{fig:profile}
}
\end{figure}

We can confirm explicitly the expectations above by solving for the bounce when we enforce minima at $\phi_\pm$.
Such minima now allow $\phi(r)$ to wait at $\phi_-$ and start rolling only after some $r_-$. This reduces the friction
and makes it possible to reach $\phi_+$ with zero velocity at some finite $r_+$. 
An example of this field profile is given in the left plot of Fig.~\ref{fig:profile}. The field takes a constant value $\phi_-$ inside an inner radius $r_-$, and reaches the false vacuum value $\phi_+$
at a finite outer radius $r_+$, with a non-trivial transition in a wall
region between both radii. The analytical solution is
\be
\Phi(r)=\left\{
\begin{matrix}
\phi_-\, , & r<r_- \\
&\\
{\displaystyle
\frac{r_\phi}{r}
k^{1/2}\ \mathrm{sn}\left[\mathrm{sn}^{-1}\left[k^{-1/2},-k^2\right]+\frac{\log(r/r_-)}{\sqrt{k^2-1}},-k^2\right]}\, ,& r_- < r < r_+\\
&\\
\phi_+\, , & r>r_+
\end{matrix}
\right.
\label{pseudobounce}
\ee
where $\mathrm{sn}(z,-k^2)$ is the Jacobi Elliptic sine function, with
\be
k^2\equiv \frac{\sqrt{1+\lambda^2r_\phi^4}+1}{\sqrt{1+\lambda^2r_\phi^4}-1}\ ,
\ee
and
\be
r_\phi \equiv r_-\phi_-=-r_+\phi_+\ .
\label{rphi}
\ee
Note that the last equality is non trivial and leads to the simple relation $r_-/r_+=-\phi_+/\phi_-$.
(As a consistency check, the standard Fubini case with $\phi_+=0, \phi_-\neq 0$ has $r_-=0$ and $r_+=\infty$,
while the thin-wall limit has $r_-\simeq r_+$).

To make the solution complete we should find a relation between $r_-/r_+$ and $r_\phi$, which is obtained
by requiring $\Phi(r_+)=\phi_+$. Using the periodicity properties of $\mathrm{sn}$ [in particular $\mathrm{sn}(x+2K(-k^2),-k^2)=-\mathrm{sn}(x,-k^2)$] we get
\be
\frac{r_-}{r_+}=\exp\left[-2 K(-k^2)\sqrt{k^2-1}\right]\ .
\label{SE}
\ee
This relation is shown in the right plot of Fig.~\ref{fig:profile}. We see that the thin-wall limit, 
$r_-/r_+\rightarrow 1$, corresponds to $r_\phi\gg 1$ while the Fubini case ($\phi_+=0$) corresponds
to $r_\phi\rightarrow 0$. 

\begin{figure}[t!]
\begin{center}
\includegraphics[width=0.6\textwidth]{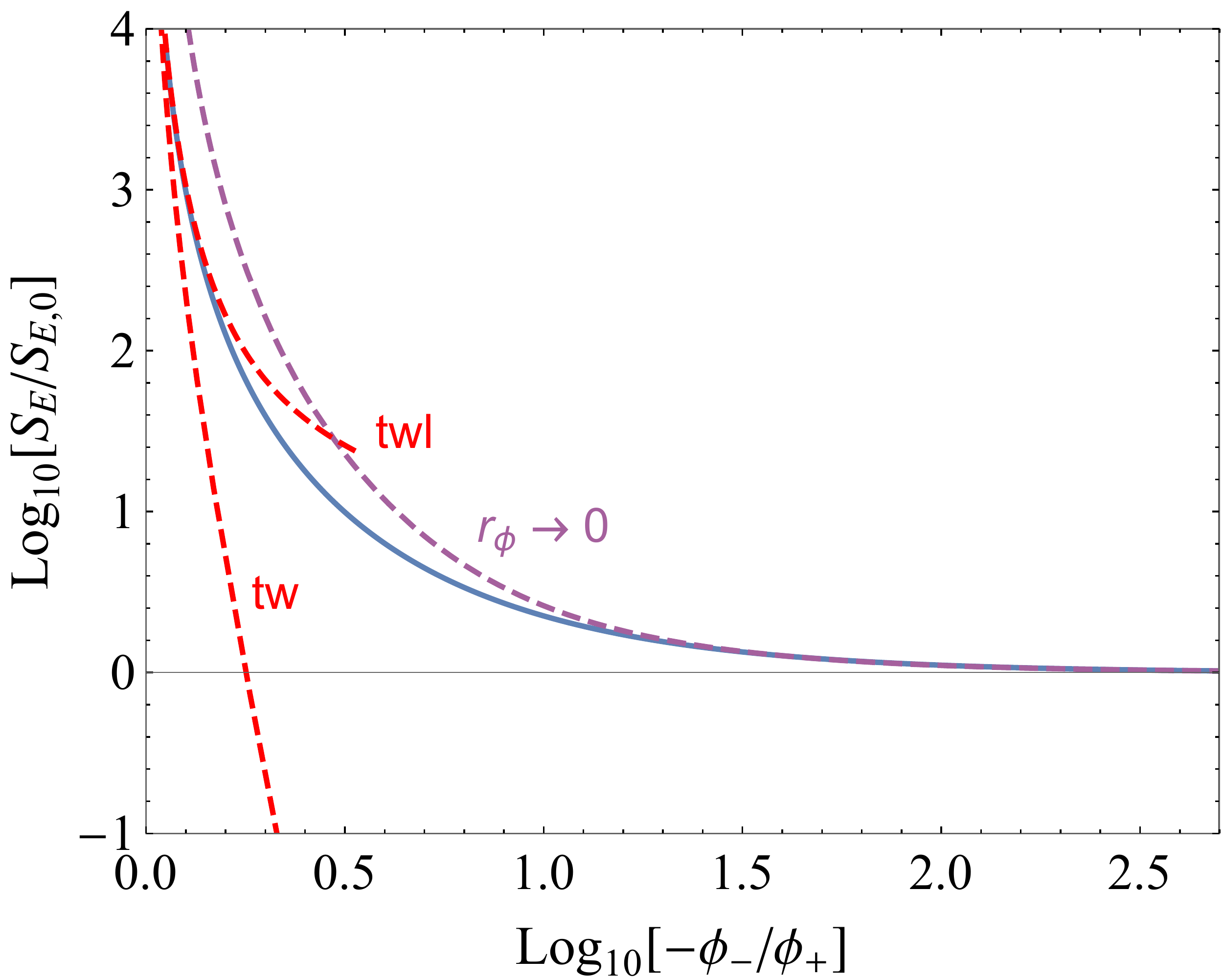}
\end{center}
\caption{Solid blue line: Euclidean action for the tunneling $\phi_+\rightarrow\phi_-$ in the potential of Fig.~\ref{fig:V}
as a function of the ratio $-\phi_-/\phi_+$ as given by Eq.~(\ref{SE}) and normalized with respect to the Fubini action $S_{E,0}=8\pi^2/(3\lambda)$. Dashed curves show approximations for $-\phi_+/\phi_-\rightarrow 0$ (or equivalently $r_\phi\rightarrow 0$, violet) and for  $-\phi_-/\phi_+\rightarrow 1$ (thin-wall limit, red).
\label{fig:SE}
}
\end{figure}

The field profile of (\ref{pseudobounce}) would be a proper bounce if the potential did have a true minimum in $\phi_-$, so we will call it ``pseudo-bounce''. This fact is at the basis of many good properties
of such pseudo-bounces, as we show later on.
It is intuitively clear that vacuum fluctuations are able to produce such configuration whether there is a true vacuum at $\phi_-$ or not,
so we argue that this field configuration can still mediate vacuum decay when the potential is unbounded. Assuming this is the case (and we will put this on solid ground in the next section) the rate would be given as usual by the Euclidean action integral, that can be performed analytically, and reads
\be
S_E(\phi_+\rightarrow \phi_-) = \frac{4\pi^2}{3\lambda\sqrt{k^2-1}}\left[2E(-k^2)-\frac{k^2-2}{k^2-1}K(-k^2)\right]\ ,
\label{SEphi4}
\ee
where $E(-k^2)$ is the complete elliptic integral of the second kind
\be
E(-k^2)=\int_0^{\pi/2}\sqrt{1+k^2\sin^2\theta}\, d\theta\ ,
\label{Ek}
\ee 
and $K(-k^2)$ has already appeared in (\ref{Kk}).

This action (\ref{SEphi4}) is shown in Fig.~\ref{fig:SE} [in units of $S_{E,0}=8\pi^2/(3\lambda)$] as a function of the ratio
$-\phi_-/\phi_+$, showing the anticipated behaviour. 
For comparison, the red dashed lines also show the thin-wall approximation of Eq.~(\ref{SEtw}) [with the wall tension calculated 
as in (\ref{sigma}), (label tw), or as explained in footnote \ref{twl} (label twl)]. The violet dashed line corresponds instead to the expansion of (\ref{SE}) for small $r_\phi$:
\be
S_E\simeq \frac{8\pi^2}{3\lambda} + \frac{1}{2}\pi^2\lambda r_\phi^4 + {\cal O}(r_\phi^8)\ .
\ee

It can also be shown that the profile (\ref{pseudobounce}) reproduces the Fubini instanton in the limit $r_\phi\rightarrow 0$ (with $r_-\rightarrow 0$). For this purpose it is convenient to rewrite (\ref{pseudobounce}) in the alternative form (for $r_-<r<r_+$)
\be
\Phi(r)=\frac{r_\phi\left(\sqrt{1-s^2}-s\sqrt{2\kappa-1}\sqrt{1-\kappa s^2}\right)}{r\left[1-s^2\left(\kappa-\sqrt{(1-\kappa)\kappa}\right)\right]}\ ,
\ee
where
\be
s\equiv\mathrm{sn}(\log(r_-/r)/\sqrt{2\kappa-1},\kappa)\ ,\quad\quad
\kappa\equiv \frac{k^2}{k^2+1}\ .
\ee
The limit $r_\phi\rightarrow 0$ corresponds to $\kappa\rightarrow 1$, in which case we can use
the expansion
\be
sn(u,\kappa) = \tanh(u)+\frac{1}{4}(1-\kappa)\left[\sinh(u)\cosh(u)-u\right]\mathrm{sech}(u)^2+{\cal O}(1-\kappa)^2\ ,
\ee
to arrive at 
\be
\lim_{r_\phi\rightarrow 0}\frac{1}{\Phi(r)}=\frac{1}{\phi_-}+\frac{1}{8}\lambda\phi_-^2r^2\ ,
\ee
which reproduces the Fubini instanton (\ref{Fubini}) with $\phi_0=\phi_-$.

What is special then about this class of potentials is that the tunneling action $S_E(\phi_+\rightarrow \phi_-)$  is a monotonically decreasing function of $\phi_-$ with the smallest value reached only asymptotically at $\phi_-\rightarrow\infty$. This explains why there is no bounce solution: the minimum of the action is pushed away to infinity. In other words, $\phi_+\neq 0$ breaks explicitly the scale invariance and lifts the flat direction of Fubini bounces into a valley in configuration space. We can still use $\phi_-$ as parameter along the bottom of this valley. The action along it decreases towards the 
scale-invariant value, achieved only asymptotically, when the 
scale breaking parameter $-\phi_+/\phi_-\rightarrow 0$. This 
shows that the decay of the $\phi_+$ vacuum is dominated by small size instantons and is therefore sensitive to ultraviolet effects that
might modify the potential at large field values. Renormalization effects can also play an important role in modifying the shape of the valley bottom. For the final calculation of the decay rate one has to integrate the differential decay rate along this valley using the collective coordinate method and deal with possible divergences \cite{OldValley2,NewValley,Matthew}.

\section{Minkowski Approach\label{sec:wkb}}

The Euclidean bounce approach to calculating tunneling actions is
ultimately justified by the WKB approach  in Minkowski space \cite{BBW,BC}. In this section we use this approach to show that pseudo-bounce field configurations like (\ref{pseudobounce}) can indeed mediate vacuum decay even though they are not proper bounces. Before proving this, we first show that these configurations share one of the key properties of the bounce: the slice of the bounce at zero Euclidean time  ($\tau=0$) gives a bubble configuration of zero energy in 3-dimensional real space. This critical bubble is the (most likely) end product of the tunneling process
out of the false vacuum. These are the bubbles that expand after being nucleated and eat away the false vacuum. 

Let us check that the $\tau=0$ slice of the Euclidean pseudo-bounce solution has indeed zero-energy. The total energy is
given by the integral
\be
E_B \equiv 4\pi\int_0^\infty dr\, r^2\left[\frac12 \left(\frac{d\Phi}{dr}\right)^2+V(\Phi)-V_+\right]\ ,
\ee
where now $r=\sqrt{\vec{x}^2}$ and $V_\pm\equiv V(\phi_\pm)$. 
We can split the integral in three pieces: the bulk ($r<r_-$), the wall ($r_-<r<r_+$) and the outside ($r>r_+$). The bulk piece is trivial and gives the negative contribution
\be
E_{B,B}=-4\pi\left.\delta V\frac{r^3}{3}\right|_0^{r_-}=-4\pi\delta V\frac{r_-^3}{3}\ ,
\ee
where $\delta V\equiv V_+-V_-$.
The wall contribution can be calculated most easily by using integration by parts and the equation of motion of
$\Phi(r)$, Eq.~(\ref{EoM4}), as follows:
\bea
\int_{r_-}^{r_+} dr\, r^2(V-V_+) &=&\left.\frac{r^3}{3} \left[V(\Phi(r))-V_+\right] \right|_{r_-}^{r_+}
-\int_{r_-}^{r_+}\frac{r^3}{3} V' \dot\Phi dr\nonumber\\
&=&\frac{r_-^3}{3}\delta V -\int_{r_-}^{r_+} \frac{r^3}{3}
\left[\ddot\Phi+\frac{3}{r}\dot \Phi\right] \dot \Phi dr
\nonumber\\
&=& 
\frac{r_-^3}{3} \delta V-\frac16 r^3 \left.\dot \Phi^2\right|_{r_-}^{r_+}-\frac12
\int_{r_-}^{r_+} \dot \Phi^2 r^2 dr\ ,
\eea
where $V'=dV/d\Phi$ and $\dot\Phi=d\Phi/dr$. Using $\dot\Phi(r_\pm)=0$ we then obtain that the wall contributes to the bubble energy the positive amount
\be
E_{B,W}=4\pi\int_{r_-}^{r_+} dr\, r^2\left[\frac12 \left(\frac{d\Phi}{dr}\right)^2+V(\Phi)-V(\phi_+)\right] =  4\pi\frac{r_-^3}{3}\delta V\ .
\ee
The energy contribution from the outside piece trivially vanishes, $E_{B,O}=0$, so adding all pieces together
we get
\be
E_B=E_{B,B}+E_{B,W}+E_{B,O}=0\ .
\ee
This nice property ultimately follows from the fact that the pseudo-bounce configuration is in fact a proper bounce of the modified
potential with a true minimum at $\phi_-$, so it inherits some good properties of proper bounces.

The connection between the Euclidean approach of Coleman and the 
WKB Minkowskian formulation to describe false vacuum decay in field theory is very clearly explained in \cite{BC}, which we follow closely below. The vacuum decay is a process of quantum tunneling between the initial vacuum configuration $\phi_+=\phi(\vec x,\alpha(t_1))$ (with $V_+=0$) and a zero-energy field configuration
$\phi(\vec x,\alpha(t_2))$  containing a bubble inside which the field probes the regions where the potential is negative.
These two configurations are separated by an energy barrier whose shape depends on the particular path in configuration
space, parametrized by $\alpha(t)$, that connects them.
The original Minkowskian action for the scalar field
\be
S=\int d^3\vec x\, dt\left[\left(\frac{d\phi}{dt}\right)^2-\frac12(\vec\nabla \phi)^2-V(\phi)+V_+\right]\ ,
\ee
restricted to a tunneling path $\phi_\alpha\equiv\phi(\vec x,\alpha(t))$, leads to
\be
S=\int_{t_1}^{t_2} dt \left[\frac12 m(\alpha)\left(\frac{d\alpha}{dt}\right)^2-{\cal V}(\alpha)\right]\ ,
\ee
with
\be
m(\alpha)\equiv\int d^3\vec x \left(\frac{d\phi_\alpha}{d\alpha}\right)^2\ ,\quad
{\cal V}(\alpha)\equiv\int d^3\vec x \left[\frac12(\vec\nabla \phi_\alpha)^2+V(\phi_\alpha)-V_+\right]\ ,
\ee
thus reducing the problem to a one-dimension quantum mechanical one. The tunneling exponent for decay along this path is then given by the usual WKB expression
\be
S_{WKB}=2\int_{\alpha_1}^{\alpha_2}\sqrt{2m(\alpha){\cal V}(\alpha)}\ ,
\ee
with $\alpha_i=\alpha(t_i)$.
Decay proceeds most likely along the path that minimizes this tunneling action (dubbed the ``most probable escape route'').
It can be shown \cite{BC} that $S_{WKB}$  agrees with the Euclidean action result taking Euclidean time to satisfy
\be
\frac{d\tau}{d\alpha} = \sqrt{\frac{m(\alpha)}{2{\cal V}(\alpha)}}\ ,
\ee
and is minimized precisely for a path related to the Euclidean bounce solution $\phi_B(r)$ by
\be
\phi_\alpha = \phi(\vec x,\alpha(t))=\phi_B(\sqrt{\vec x^2+\tau^2})\ ,
\ee
making $O(4)$ invariance manifest.

\begin{figure}
\begin{center}
\includegraphics[width=0.6\textwidth]{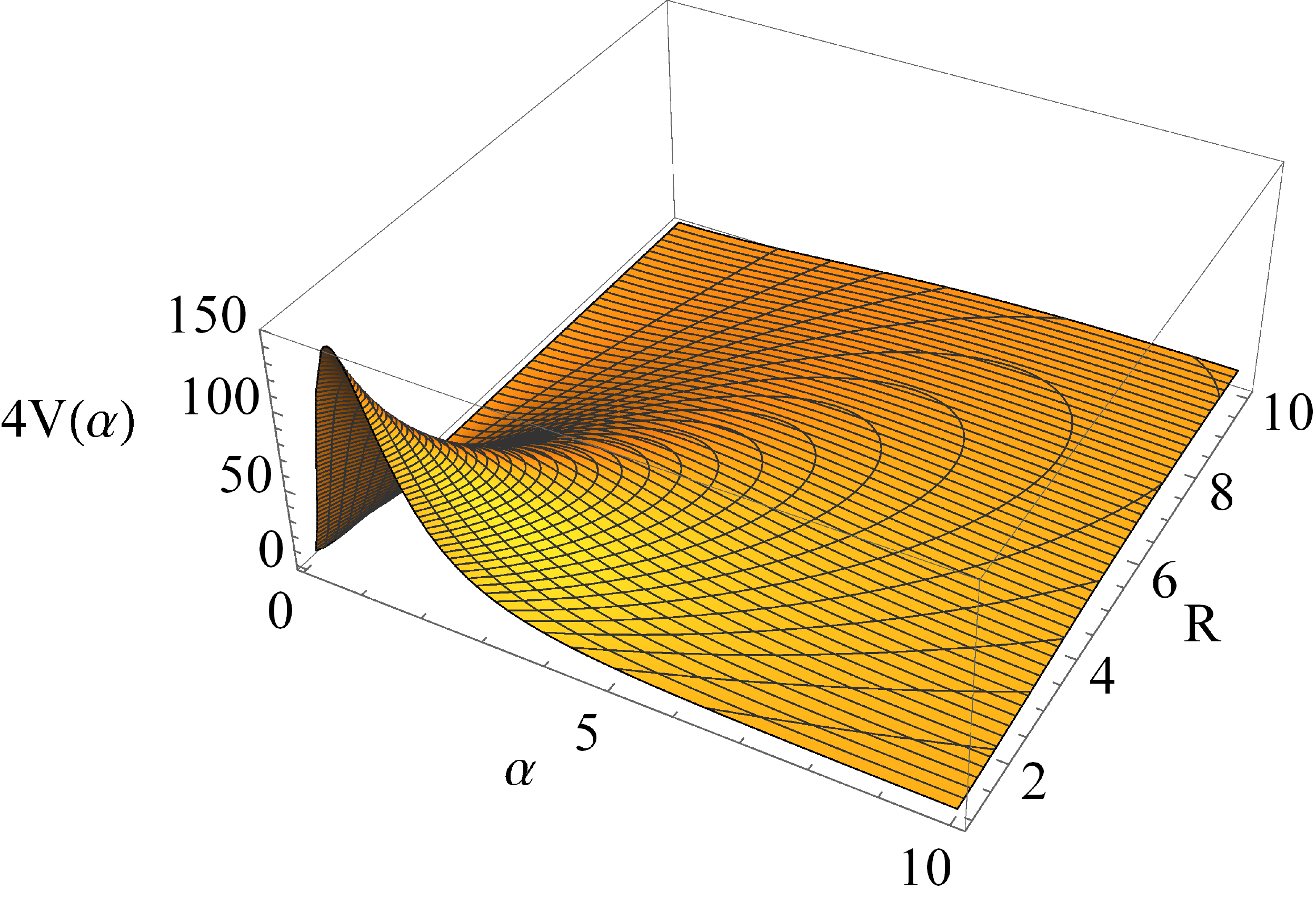}
\end{center}
\caption{Energy barrier in configuration space [spanned by the parameters $\alpha$ and $R$ of the path in (\ref{path})] for  tunneling out of $\phi_+=0$ with $V(\phi)=-\lambda\phi^4/4$. (For this plot $\lambda=0.1$).
\label{fig:EBl4}
}
\end{figure}

To illustrate this approach consider the simple case \cite{Gino} with $V(\phi)=-\lambda\phi^4/4$ and take the trajectory in field space to be
defined by 
\be
\phi_\alpha(r,\alpha)=\frac{\phi_0}{1+(r^2+\alpha^2)/R^2}\ ,\quad  \mathrm{with} \;\; \frac{1}{R^2}=\frac{\lambda\phi_0^2}{8}\ ,
\label{path}
\ee
with $r=\sqrt{\vec x^2}$.
For $\alpha=\infty$ we get the false vacuum, $\phi_\alpha\rightarrow\phi_+=0$, while $\alpha=0$ gives the $\tau=0$ slice of the Fubini instanton
profile.
It is straightforward to get
\be
m(\alpha)=2{\cal V}(\alpha)=\frac{4\pi^2\alpha^2R^2}{\lambda(R^2+\alpha^2)^{5/2}}\ ,
\ee
so that $\alpha=\tau$, and then
\be
S_{WKB}=4\int_0^\infty {\cal V}(\alpha)d\alpha=\left.\frac{8\pi^2\alpha^3}{3\lambda(R^2+\alpha^2)^{3/2}}\right|_0^\infty=\frac{8\pi^2}{3\lambda}\ ,
\label{WKBl4}
\ee
precisely the value obtained via the Euclidean approach.
Figure~\ref{fig:EBl4} shows $4{\cal V}(\alpha)$ for different values of $R$ (or $\phi_0$). This is the true energy barrier under which vacuum decay tunneling takes place. Although the height of the barrier changes with $R$, the area below the barrier, that determines the tunneling exponent as in (\ref{WKBl4}), remains constant, as obtained explicitly above.

\begin{figure}[t!]
\includegraphics[width=0.45\textwidth]{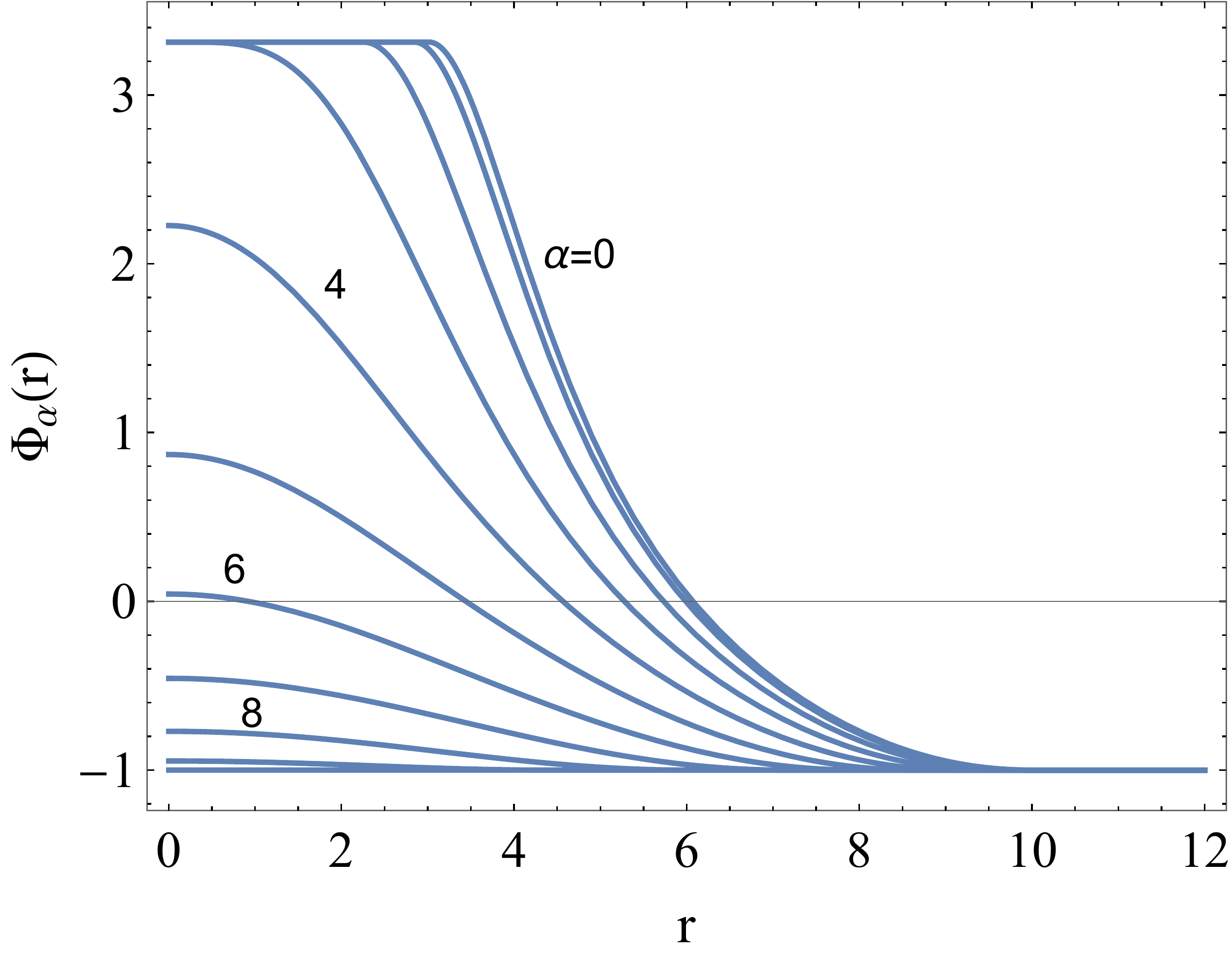}\,\,
\includegraphics[width=0.55\textwidth]{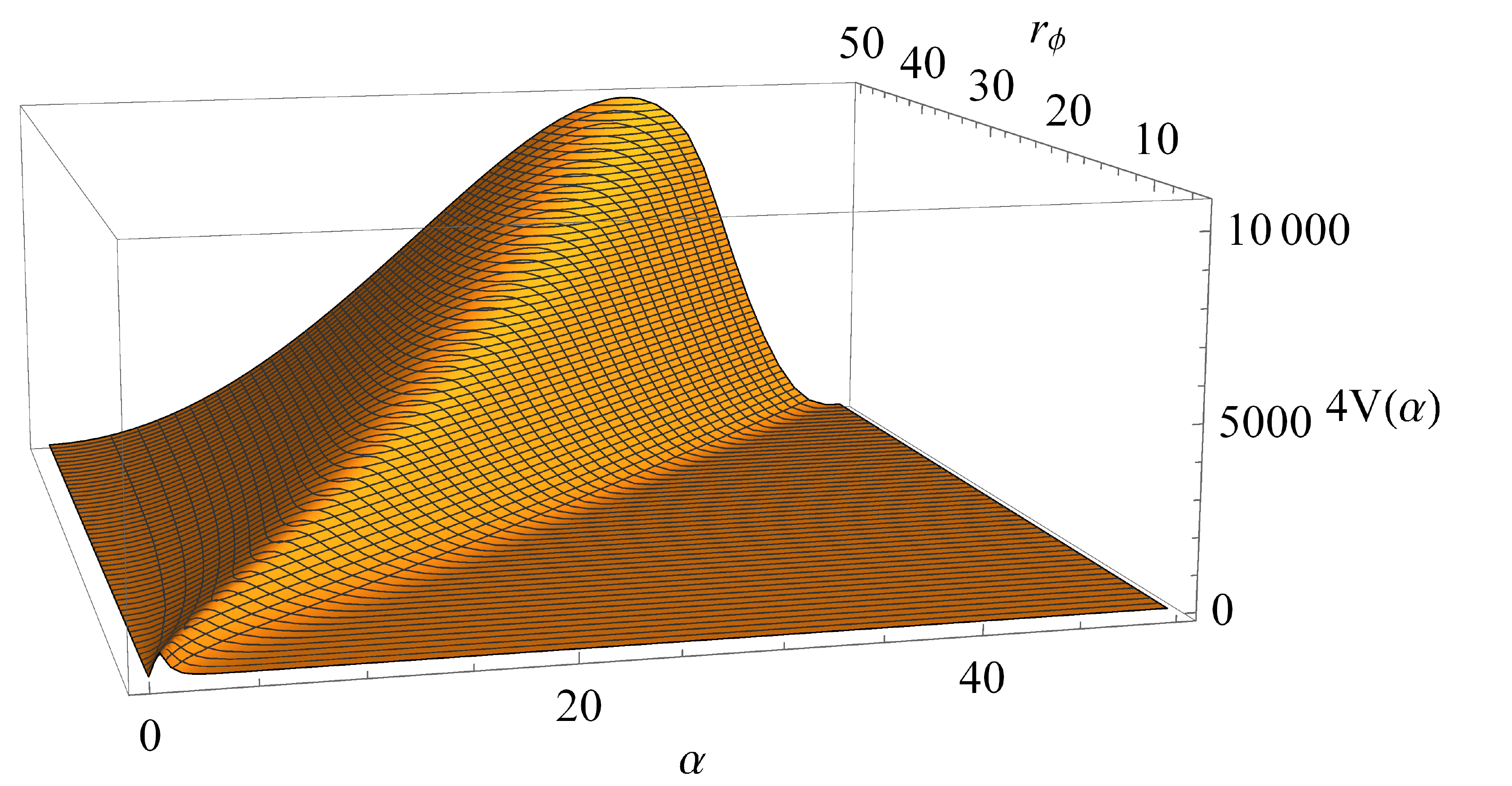}
\caption{Left: Snapshots of the tunneling path~(\ref{path2})
for the indicated values of $\alpha$. For $\alpha=0$ we get the tunneling bubble configuration while $\alpha=10$ gives the false vacuum $\phi_+=-1$. Right: Energy barrier in configuration space [spanned by the parameters $\alpha$ and $r_\phi$ of the path (\ref{path2})] for tunneling out of $\phi_+=-1$ with  $V(\phi)=-\lambda\phi^4/4$.
We use $\lambda=0.1$, $r_+=r_\phi=10$.
\label{fig:PT}
}
\end{figure}

Let us then consider a pseudo-bounce profile like Eq.~(\ref{pseudobounce}) (which we will use for the numerical examples)
and define a path in Minkowski space simply via the replacement
$r\rightarrow \sqrt{r^2+\alpha^2}$
\be
\Phi_\alpha(r,\alpha)\equiv \Phi(\sqrt{r^2+\alpha^2})\ .
\label{path2}
\ee
Different $\alpha$ snapshots of this path for the numerical example are shown in Fig.~\ref{fig:PT}, left plot. In this case, $\alpha=0$ corresponds to the pseudo-bounce configuration and $\alpha=r_+$ corresponds to the false vacuum $\phi_+$ (taken at $\phi_+=-1$, with $r_\phi=10$, in this example).
Using $O(3)$ rotational invariance and the fact that inner and outer radii for general $\alpha$ are $r_{\alpha,\pm}\equiv \sqrt{r_\pm^2-\alpha^2}$  we can write
\bea
m(\alpha) &=& 4\pi\alpha^2\int_{r_{\alpha,-}}^{r_{\alpha,+}}  \dot\Phi_\alpha^2\, dr \ ,\nonumber\\
{\cal V}(\alpha)&=&-\frac43 \pi\delta V r_{\alpha,-}^3+ 
4\pi\int_{r_{\alpha,-}}^{r_{\alpha,+}}\left[\frac12 \dot\Phi_\alpha^2+V(\phi_\alpha)-V_+\right]\ ,
\eea
where $\dot\Phi_\alpha=d\Phi_\alpha/dr$. The differential equation satisfied by $\Phi_\alpha(r)$ in the interval $(r_{\alpha,-},r_{\alpha,+})$ can be derived directly from Eq.~(\ref{EoM4}) and is
\be
\left(1+\frac{\alpha^2}{r^2}\right)\ddot\Phi_\alpha+\frac{1}{r}\left(3-\frac{\alpha^2}{r^2}\right)\dot\Phi_\alpha=V'(\Phi_\alpha)\ .
\ee
Using this equation and integration by parts in the ${\cal V}(\alpha)$ integral above shows that $m(\alpha)=2{\cal V}(\alpha)$. Therefore
\be
S_{WKB}=\int_0^{r_+}4{\cal V}(\alpha)d\alpha\ .
\label{SWKB}
\ee
Fig.~\ref{fig:PT}, right plot, shows the integrand above in our numerical example as a function of $r_\phi$ (which ultimately determines $\phi_0$
and the size of the pseudo-bounce). As is clear from the figure, tunneling at lower values of $r_\phi$ is preferred, as was found in section~\ref{sec:lambdaphi4}. In fact, it can be checked numerically that  (\ref{SWKB}) agrees with the analytical result for the Euclidean action given in Eq.~(\ref{SEphi4}). 

The previous discussion confirms (and illustrates numerically) that the Euclidean calculation of section~\ref{sec:lambdaphi4} gives indeed the correct
action corresponding to a vacuum decay by tunneling. Moreover, 
$S_{WKB}$ minimizes the action for tunneling from $\phi_+$ towards a fixed value $\phi_-=\phi_0$ (as is clear by thinking about the modified potential with a true minimum at $\phi_-$).

\section{Tunneling potential approach\label{sec:Vt}}

The new approach proposed in
\cite{E} reformulates the calculation of tunneling actions as a simple variational problem in field space, without reference to Euclidean space or bounces. 
This tunneling potential formulation has a number of advantages
and appealing features that have been studied in detail elsewhere. It allows for a fast and precise numerical determination of the action; it can be modified to study decays by thermal fluctuations; it can be used to construct potentials that allow a fully analytical solution to the  tunneling problem \cite{E}; it can be generalized to include in a simple and compact way gravitational corrections \cite{Eg} and
it can be very useful to study efficiently vacuum decays in multi-field potentials \cite{EK} as one is searching for a minimum of the action (rather than a saddle-point, as in the Euclidean case).

For a potential $V(\phi)$ with a false vacuum
at $\phi_+$, the tunneling action is obtained as the minimum
of the functional
\be
S[V_t] \equiv 54\pi^2\int_{\phi_+}^{\phi_0}\frac{(V-V_t)^2}{-(V_t')^3}d\phi\ ,
\label{SVt}
\ee
where $V_t'=dV_t/d\phi\leq 0$. 
The Euler-Lagrange equation corresponding to minimizing this action
reads \cite{E}:
\be
(4V_t'-3V')V_t'=6(V_t-V)V_t''\ .
\label{EoMVt}
\ee
The tunneling potential $V_t(\phi)$
to be found has to satisfy the boundary conditions 
\be
V_t(\phi_+)=V(\phi_+)\ , \quad\quad V_t(\phi_0)=V(\phi_0)\ .
\label{BC}
\ee 
The correspondence between this formulation and the Euclidean one is based on
the relation $V_t=V-\dot\phi_B^2/2$, where $\phi_B$ is the Euclidean bounce. The field value
$\phi_0$ corresponds precisely to $\phi_B(0)$, see \cite{E} for details. 

An example of the
shape of the tunneling potential is given in Fig.~\ref{fig:Potential}, left plot.
The agreement between the actions calculated in both formalisms,
proven in \cite{E} for proper bounces, is straightforward after writing the Euclidean action in terms of the gradient contribution only. More precisely, one has
\be
S_E \equiv 2\pi^2\int_0^\infty dr\, r^3\left[\frac12 \left(\frac{d\Phi}{dr}\right)^2+V(\Phi)-V_+\right]=S_K+S_V \ ,
\ee
where we have split the action in a gradient contribution $S_K$ and
a potential contribution, $S_V$. Derrick's theorem \cite{Derrick} gives the relation $S_V=-S_K/2$ so that $S_E=S_K/2$.

In this section we examine how this new approach can be applied to potentials with false vacua that admit no bounce. It is in fact most natural in this approach to consider  what is the minimal value of the action $S[V_t]$ for a fixed value of $\phi_0$ first (we call such $V_t$ ``restricted'' tunneling potential) and then vary $\phi_0$ to find the absolute minimum of the action. From the discussion in previous sections
it is clear what outcome to expect at least for the $V=-\lambda\phi^4/4$ potential of section~\ref{sec:lambdaphi4}: $S$ will have a runaway minimum at $\phi_0\rightarrow \infty$. Before showing that,
we first check that $S[V_t]$ does reproduce the correct
value we found via the Euclidean or Minkowskian approaches.

The agreement of $S[V_t]$ and the Euclidean action $S_E$ is based on the relations $V_t=V-\dot\phi_B^2/2$ and $S_E=S_K/2$, as we have mentioned above. Let us check this last relation
for the pseudo-bounce field configuration.
We again split the action integral in  bulk ($r<r_-$),  wall ($r_-<r<r_+$) and  outside ($r>r_+$)
contributions. The bulk piece gives 
\be
S_{E,B}=-\pi^2\left.\delta V\frac{r^4}{2}\right|_0^{r_-}=-\pi^2\delta V\frac{r_-^4}{2}\ .
\ee
For the wall contribution we use again integration by parts and the equation of motion to get:
\be
\int_{r_-}^{r_+} dr\, r^3(V-V_+) =
\frac{r_-^4}{4} \delta V-\frac14
\int_{r_-}^{r_+} \dot \Phi^2 r^3 dr\ .
\ee
and then obtain for the wall contribution to the Euclidean action
\be
S_{E,W}=2\pi^2\int_{r_-}^{r_+} dr\, r^3\left[\frac12 \dot\Phi^2+V(\Phi)-V_+\right] =  
\pi^2\frac{r_-^4}{4}\delta V+ \frac{\pi^2}{2}\int_{r_-}^{r_+} dr\, r^3\dot\Phi^2\ .
\ee
The action contribution from the outside piece trivially vanishes, $S_{E,O}=0$, so adding all pieces together
we get
\be
S_E=S_{E,B}+S_{E,W}+S_{E,O}=\frac12 S_K\ ,
\ee
precisely as in the bounce case. We see once again that the pseudo-bounce has some of the good properties of a proper bounce, which  is due to the fact that the pseudo-bounce would be a true bounce for a modified potential.

While a true Euclidean bounce corresponds, in the formulation of the tunneling potential, to a $V_t$ that gives the absolute minimum for $S[V_t]$, a pseudo-bounce corresponds to a $V_t$ that minimizes
 $S[V_t]$ restricted to those $V_t$'s that end at a fixed $\phi_0$.
It is interesting that the boundary conditions on $V_t'(\phi_0)$
are different in both cases. While the $V_t$ that gives the absolute action minimum satisfies $V_t'(\phi_0)=3V'(\phi_0)/4$ \cite{E}, the ``restricted'' $V_t$ satisfies instead $V_t'(\phi_0)=0$. Both cases are consistent with Eq.~(\ref{EoMVt}). In section~\ref{sec:analyticVt}
we will present some potentials with restricted $V_t$'s that can be studied with full analytical control.

The boundary condition $V_t'(\phi_0)=0$ for restricted tunneling potentials can be understood 
resorting again to the modified potential with a sharp minimum at $\phi_-=\phi_0$, so that $V'(\phi_0)=0$ for the modified potential. Alternatively, we can see the consistency of $V_t'(\phi_0)=0$  with the pseudo-bounce profile using the relation between the Euclidean radial coordinate and $V_t$ \cite{E}
\be
r=3\sqrt{\frac{2(V-V_t)}{(V_t')^2}}\ .
\label{r}
\ee
In the proper bounce case, generically $V_t'(\phi_0)\neq 0$ so that 
 $r\rightarrow 0$ for $\phi\rightarrow\phi_0$ [as $V_t(\phi_0)=V(\phi_0)$]. For a pseudo-bounce, instead, $r$ takes a finite value $r_-\neq 0$ when $\phi\rightarrow \phi_0$ so that (\ref{r})
requires $V_t'(\phi_0)\rightarrow 0$, with
\be
\left.\frac{V-V_t}{(V_t')^2}\right|_{\phi_0}=\frac{r_-^2}{18}\ .
\ee

There is a direct link between the fact that $r_-\neq 0$ for pseudo-bounces and the fact that they do not minimize the tunneling action.
From (\ref{SVt}) we get, using integration by parts,
\bea
\frac{dS}{d\phi_0} &= &54\pi^2\left\{\left.\frac{(V-V_t)^2}{-(V_t')^3}
\right|_{\phi_0}+
3\left.\frac{(V-V_t)^2}{(V_t')^4}\frac{dV_t}{d\phi_0}\right|_{\phi_+}^{\phi_0}
\right.
\nonumber\\
&+&\left.
2\int_{\phi_+}^{\phi_0}\frac{(V-V_t)}{(V_t')^5}\left[(4V_t'-3V')V_t'+6(V-V_t)V_t''\right]\frac{dV_t}{d\phi_0}d\phi
\right\}\ .
\eea
The first term in the right hand side vanishes as it is proportional to $r_-^2 V_t'(\phi_0)=0$. The third term vanishes because the integrand is zero due to the Euler-Lagrange equation (\ref{EoMVt}). The only term surviving is the second (boundary) term, that can be rewritten as 
\be
\frac{dS}{d\phi_0} =\frac{\pi^2}{2}r_-^4 \left.\frac{dV_t}{d\phi_0}\right|^{\phi_0}_{\phi_+}=\frac{\pi^2}{2}r_-^4 V'(\phi_0)\ ,
\label{dSdphi}
\ee
where we have used $dV_t/d\phi_0|_{\phi_+}=0$ and $dV_t/d\phi_0|_{\phi_0}=V'(\phi_0)$, which follow from the boundary conditions on $V_t$, Eq.~(\ref{BC}). This simple relationship holds
for general pseudo-bounces and can be checked explicitly e.g.
for the action obtained in (\ref{SEphi4}) for the potential studied in section~\ref{sec:lambdaphi4}. 

By the same kind of arguments used in \cite{E}, it can be shown 
that restricted tunneling potentials minimize the tunneling action
for fixed end-point $\phi_0$, although the action can be lowered if $\phi_0$ is allowed to vary. This is precisely the behaviour expected for the bottom of the valley in configuration space after scale invariance is broken. The valley bottom is defined by the property of being a minimum for deviations along all directions except for one that traces the valley, for which the action has non-zero derivative.
Restricted $V_t$'s (or pseudo-bounces)  correspond therefore to this valley bottom and $\phi_0$ can be used to parametrize it.
Eq.~(\ref{dSdphi}) is similar to the stream-line equation used in
\cite{OldValley2} to parametrize the valley but does not involve
$d\Phi/d\phi_0$, which simplifies the equation significantly.

\section{Constrained instantons and valley methods\label{sec:civ}}

\begin{figure}[t!]
\includegraphics[width=0.5\textwidth]{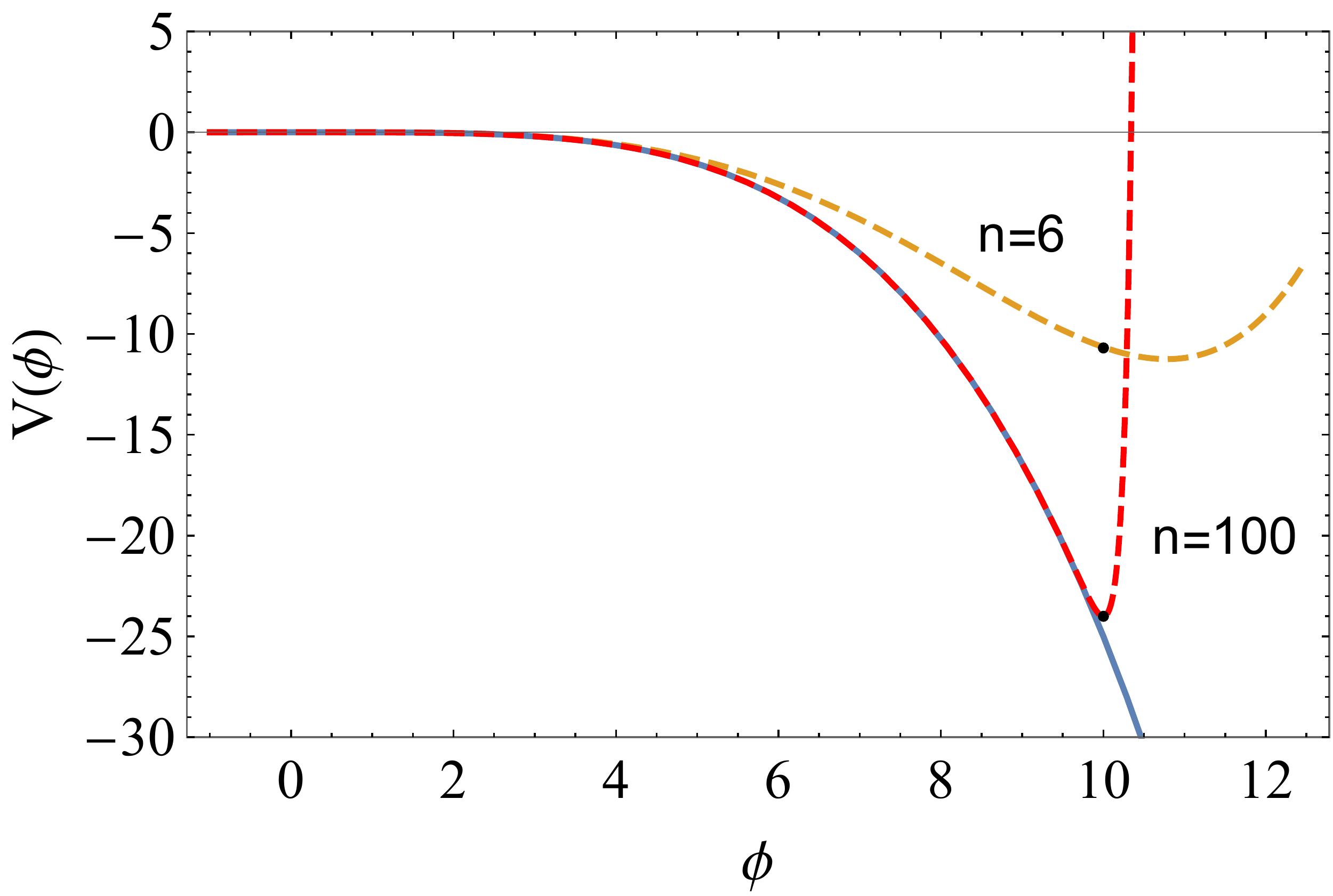}\,\,\,
\includegraphics[width=0.5\textwidth]{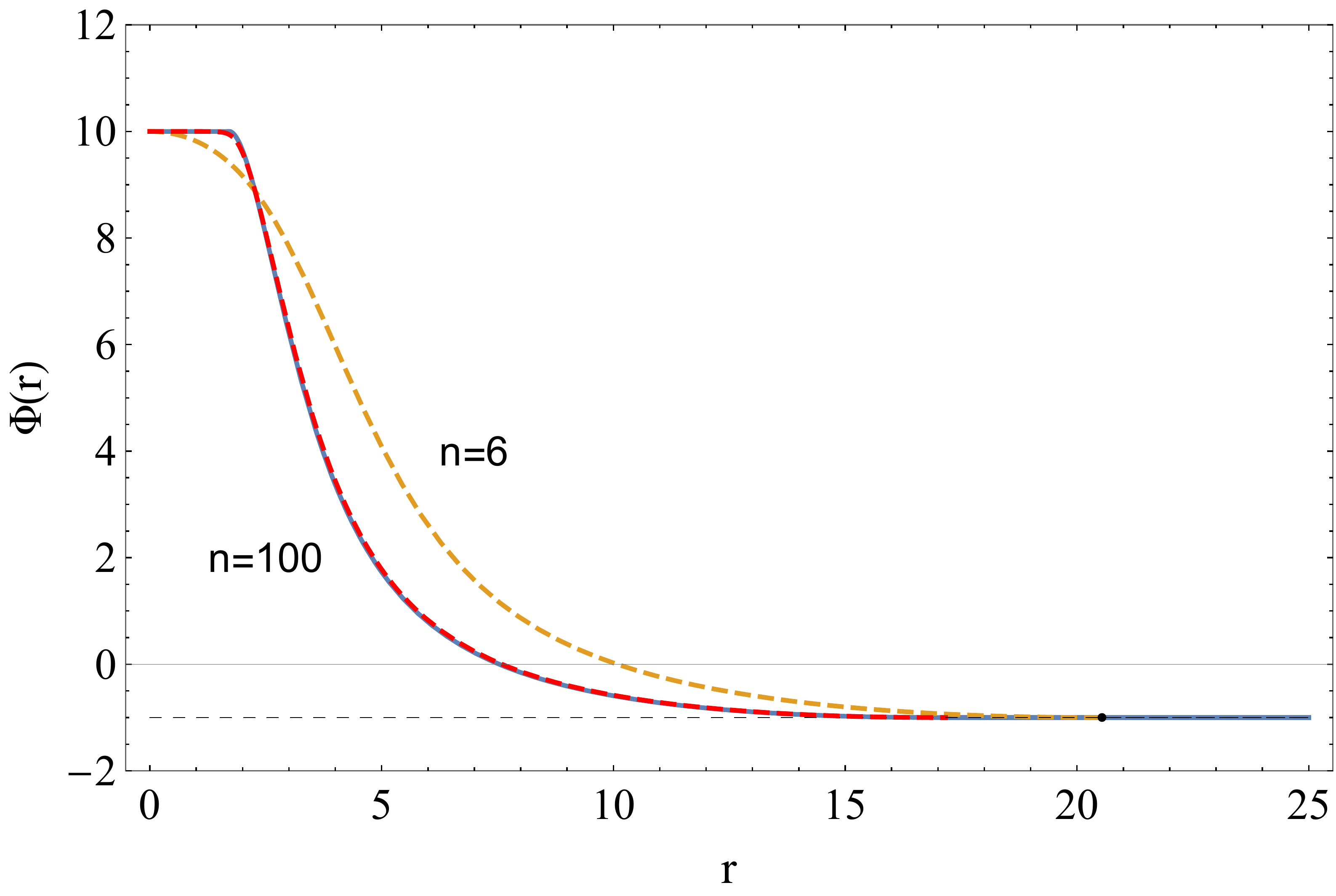}
\begin{center}
\includegraphics[width=0.5\textwidth]{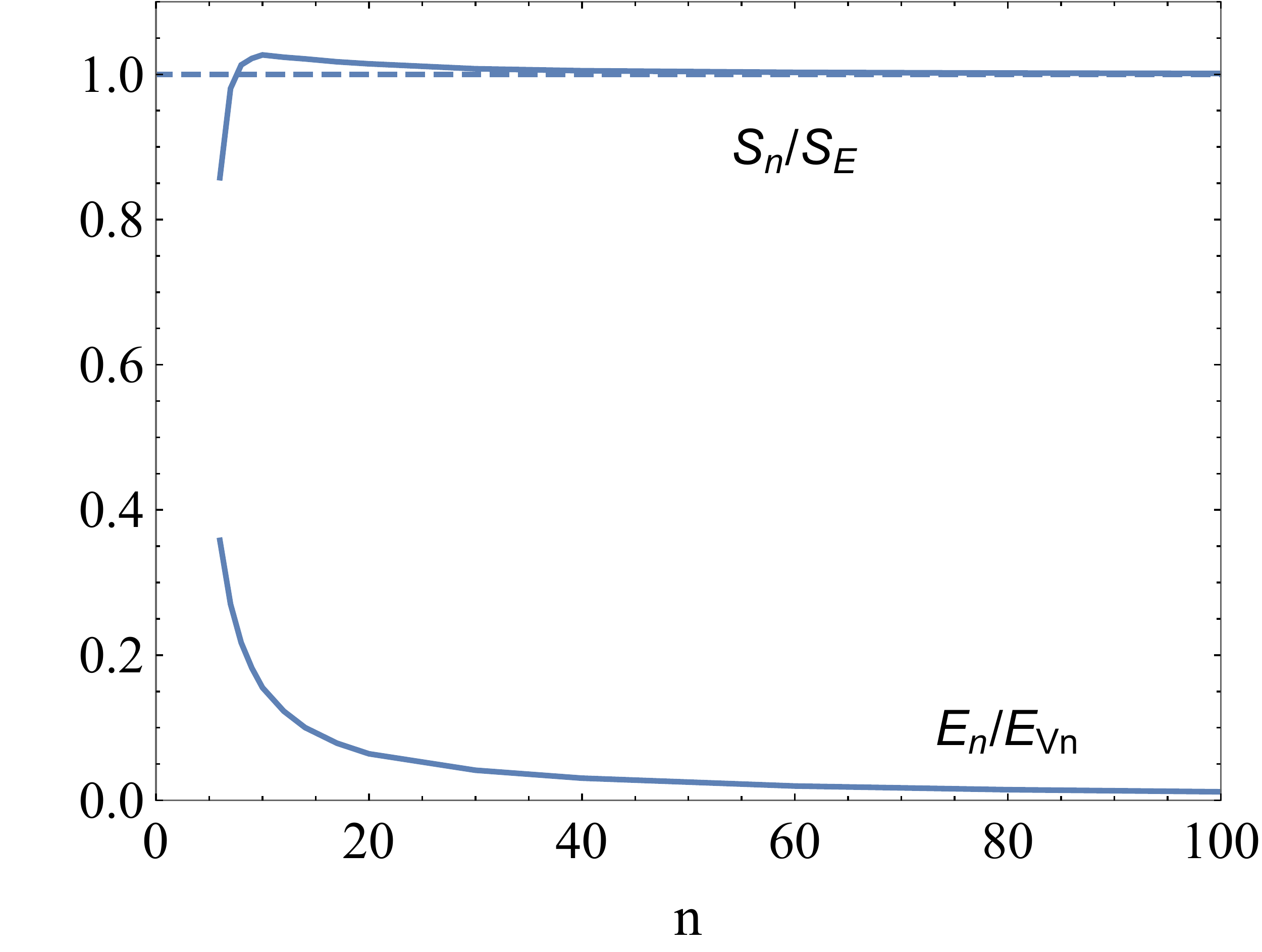}
\end{center}
\caption{Upper left: Potential $V(\phi)=-\lambda\phi^4/4$ (blue)
with $\phi_+=-1$ and $\lambda=0.01$; and two constrained potentials $V_n(\phi)=-\lambda\phi^4/4+\lambda_n\phi^n$ for $n=6,100$ (dashed lines),
with $\lambda_n$ fixed to get $\phi_0=10$ (marked by black dots).
Upper right: Profiles of pseudo-bounce (blue) and constrained bounces for $n=6,100$ corresponding to the potentials on the left plot. 
Lower: As a function of the exponent $n$ for the potentials $V_n(\phi)$, action $S_n$ of the constrained bouncess normalized to the Euclidean action of the pseudobounce $S_E$ (upper curve).
Total energy over potential energy for the slice of zero Euclidean time of the constrained bounces (lower curve).
\label{fig:constrained}
}
\end{figure}

In previous sections we have found tunneling profiles in three different formulations of the tunneling problem (Euclidean, Minkowskian and tunneling potential) for false vacua decaying without a proper bounce. Now we compare this approach with alternative solutions that were discussed in previous literature. 

The constrained instanton approach \cite{Affleck} manages to
recover a bounce by imposing a constraint on the field profile.
A typical constraint is of the form
\be
\int d^4x\,  \phi^n = C\phi_0^{n-4}\ .
\ee
By using a Lagrange multiplier, one then extremizes the new action
\be
S_n = \int d^4x\left[\frac12 (\partial_\mu\phi)^2 + V(\phi)-V_+ +\lambda_n \phi^n \right]-\lambda_n C\phi_0^{n-4}\ .
\ee
The new term added to the action modifies the potential and this allows a bounce: for a given value of $\phi_0$ the constant $\lambda_n$ is tuned to make $\phi(\infty)=\phi_+$.
However, such constrained bounces do not have in general the nice properties expected of a bounce (and shared by our pseudo-bounces). In particular, their $\tau=0$ slice
produces a three-dimensional profile that has nonzero energy and cannot be the end product of a tunneling event. In this respect, pseudo-bounces or restricted tunneling potentials
are better suited to describe the tunneling problem.

Nevertheless, one can make contact between the two approaches. 
For a given value of $\phi_0$, taking $n\gg 1$ one can
 arrange for the modified potential to have a sharp minimum at $\phi_0$  with the potential being arbitrarily close to the original potential for $\phi<\phi_0$. This is shown in Fig.~\ref{fig:constrained}, upper left plot, where the modified potential is $V(\phi)=-\lambda\phi^4/4+\lambda_n\phi^n$. This is basically the same trick we used to arrive at the pseudo-bounce in previous sections and we expect that the constrained instanton will approach the pseudo-bounce profile at large $n$. This expectation is realized as illustrated in Fig.~\ref{fig:constrained}, upper right plot. The $n=100$ profile (red) is almost on top of the pseudo-bounce (blue).
 We can also check that as $n$ grows, the constrained instanton action tends towards the pseudo-bounce action and has a $\tau=0$ slice configuration of zero-energy. This is shown in the lower plot of Fig.~\ref{fig:constrained},
 that shows, as a function of the exponent $n$, the ratio of the constrained instanton action $S_n$ over the pseudo-bounce action $S_E$ (tending to 1 for large $n$) and, for the $\tau=0$ slice of the constrained bounce, the ratio of its total energy over the potential energy (tending to zero for large $n$).
 We conclude
 that we can think of the pseudo-bounces as constrained instantons for which the constraint is basically $\phi(r)\leq \phi_0$ (constraint that is most naturally implemented in the tunneling potential approach).

The so-called new valley method \cite{NewValley} (see \cite{OldValley1,OldValley2,OldValley3} for previous related work) sets up two differential equations to trace the bottom of the valley along the flat direction lifted by the breaking of scale invariance. The idea of the method  is to define that valley as the most gentle direction of variation of the Euclidean action (for slices of fixed action) introducing an auxiliary field that measures by how much the Euler-Lagrange equation for the bounce is not satisfied. Without entering into the details, the resulting Euclidean profiles suffer the same shortcoming of constrained instantons of finite $n$: their $\tau=0$ slice
is not guaranteed to have zero energy. Presumably, the presence of the negative eigenvalue of the Euclidean action interferes with the successful implementation of this idea. If one implements the same idea using the new action $S[V_t]$ for the tunneling potential
given in (\ref{SVt}) (instead of the Euclidean action),
the result is trivially the same we have presented in the previous section. The reason is that the restricted tunneling potential satisfies the Euler-Lagrange equation (\ref{EoMVt}), so that the corresponding auxiliary field trivially vanishes.

\section{Potential $\bma{V(\phi)=-\frac{1}{2}m^2\phi^2}$\label{sec:m2}}

In section~\ref{sec:lambdaphi4} we analyzed tunneling in the scale invariant  potential $V=-\lambda\phi^4/4$, breaking scale invariance by fixing vacua at $\phi_\pm\neq 0$. In that case all
solutions of the Euler-Lagrange equation (\ref{EoM4}) are undershots and there is no bounce. In this section we analyze another very simple potential:
\be
V(\phi)=-\frac{1}{2}m^2\phi^2\ ,
\label{Vm2}
\ee
(with $m>0$) which is standard in the sense that it has under- and over-shots and therefore a bounce. Still one can consider the possible tunnelings not mediated by the bounce. Like we did in
section~\ref{sec:lambdaphi4}, we locate a false minimum
at $\phi_+<0$ and examine decays to some assumed
 deeper minimum at $\phi_->-\phi_+$.

The true bounce profile can be obtained analytically as 
\be
\Phi_B(r)=\frac{2\phi_-^B}{m r}J_1(m r)\ ,
\label{bouncem2}
\ee
where $J_\alpha(x)$ is the Bessel function of the first kind. The bounce starts at $\phi_-^B$ for $r=r_-^B=0$ and reaches the false vacuum $\phi_+$ at a finite radius $r_+^B$, where $\dot\Phi_B( r_+^B)=0$. As $\dot\Phi_B\propto J_2(mr)$, $r_+$ is determined by the first zero of $J_2(x)$ to be  $m r_+^B\simeq 5.13562$. From the bounce expression (\ref{bouncem2}) it follows that
\be
-\frac{\phi_+}{\phi_-^B}= \frac{-2J_1(m r_+^B)}{m r_+^B}\simeq 0.132279\ ,
\ee
which fixes $\phi_-^B$ for a given $\phi_+$.
The corresponding bounce action is
\be
S_B=\frac{\pi^2}{4}\phi_+^2m^2(r_+^B)^4\simeq \left(41.4292\,\frac{\phi_+}{m}\right)^2 .
\label{SB}
\ee

\begin{figure}[t!]
\includegraphics[width=0.5\textwidth]{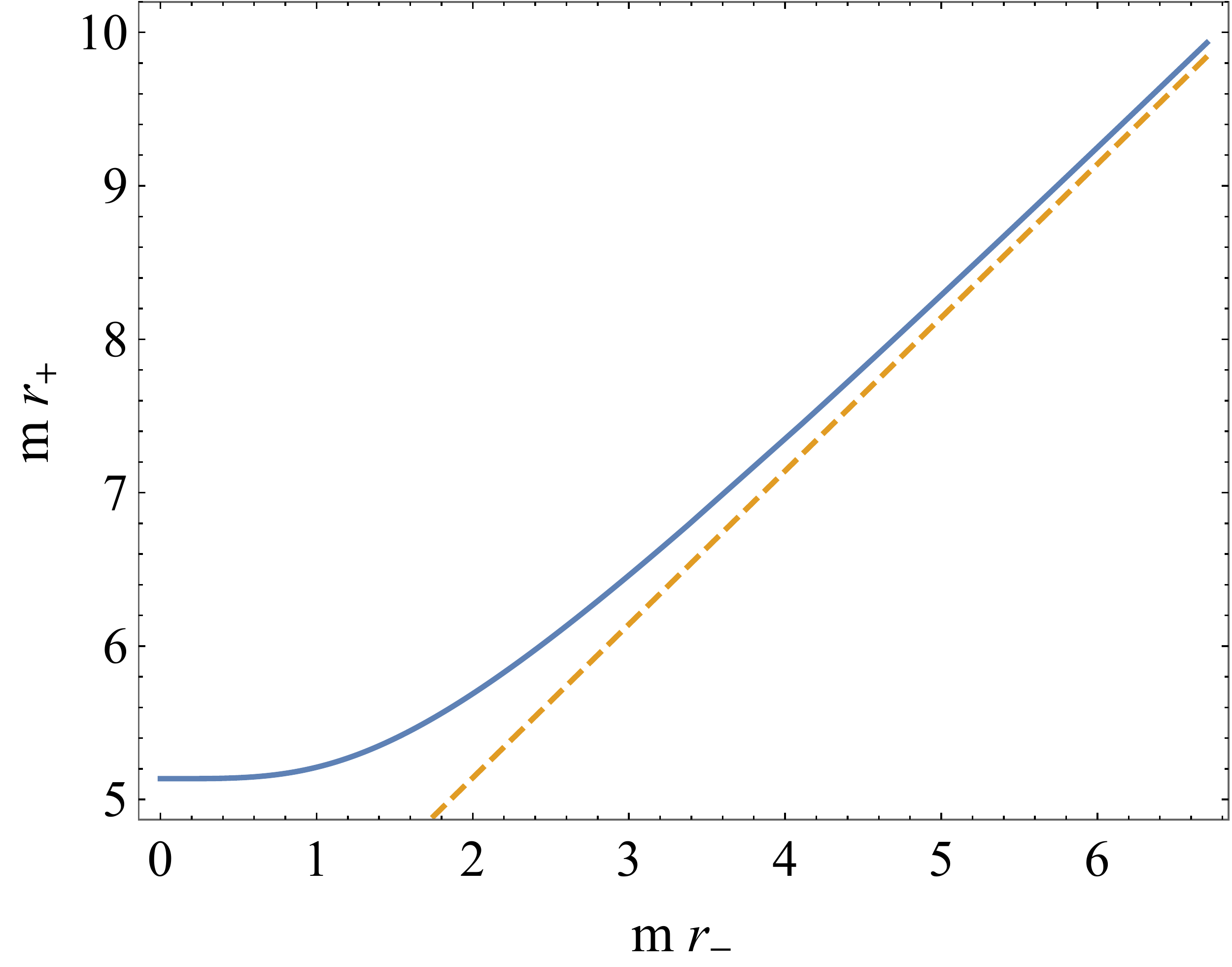}\,\,\,\includegraphics[width=0.5\textwidth]{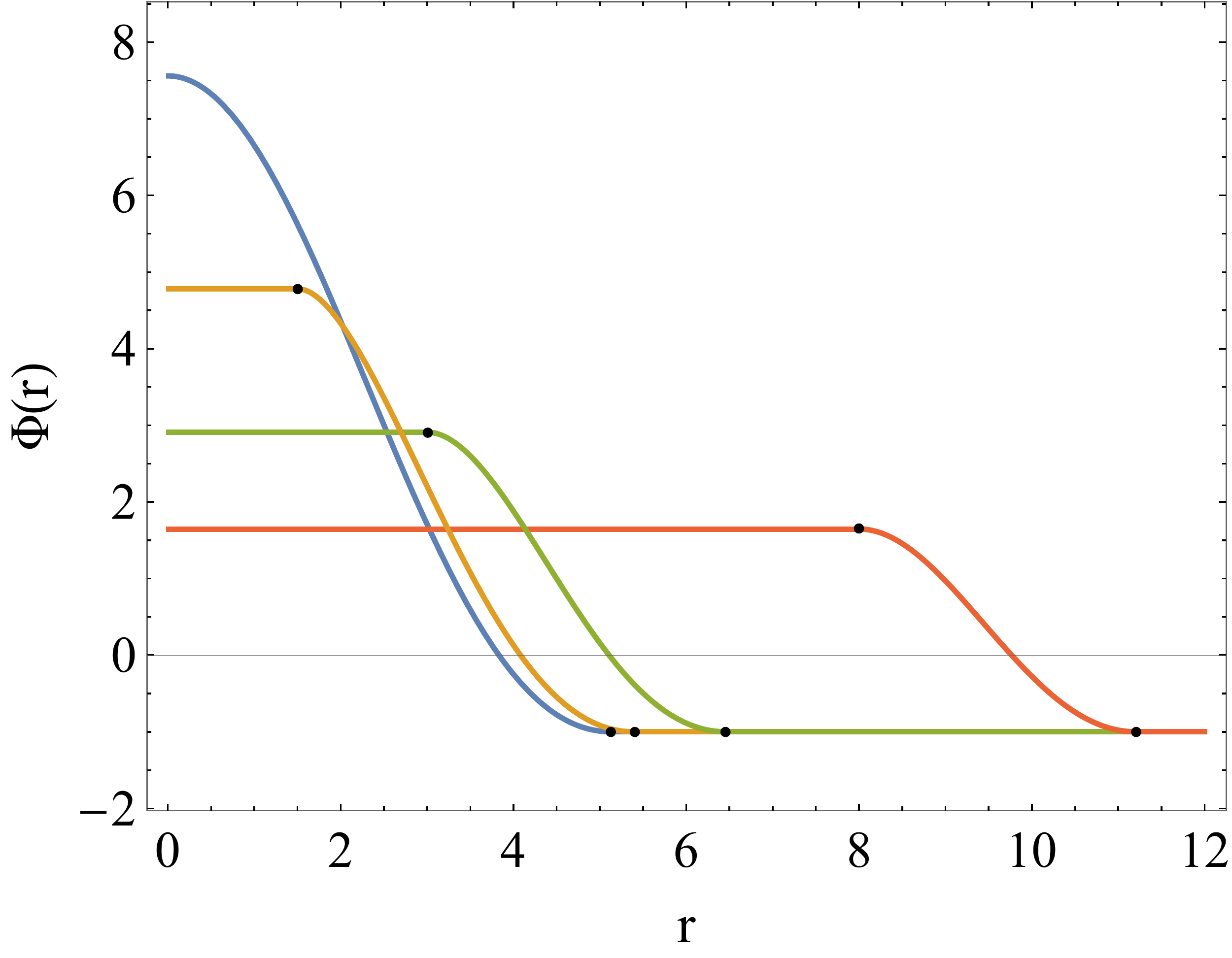}
\includegraphics[width=0.5\textwidth,height=0.4\textwidth]{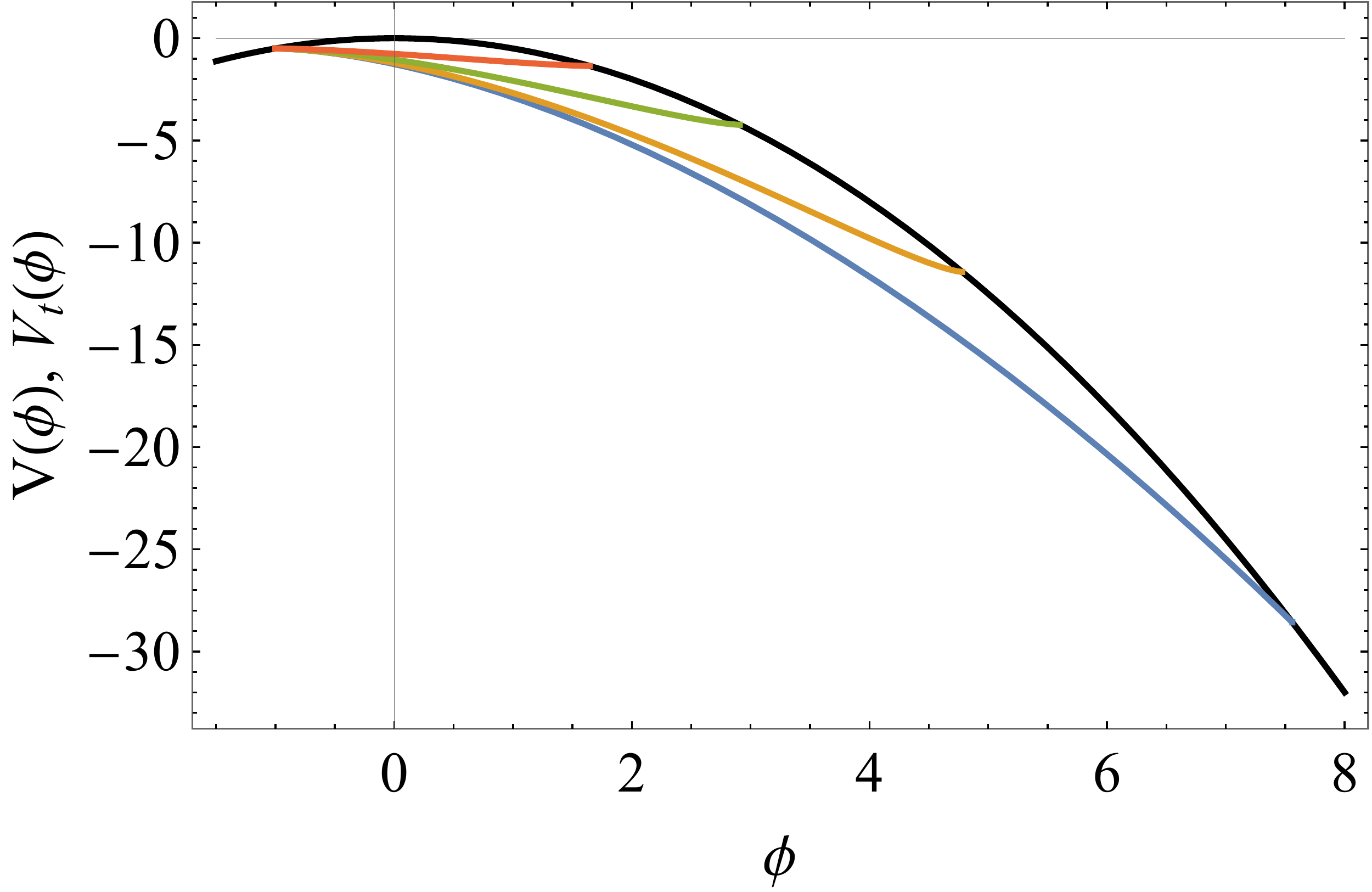}\,\,\,
\includegraphics[width=0.5\textwidth]{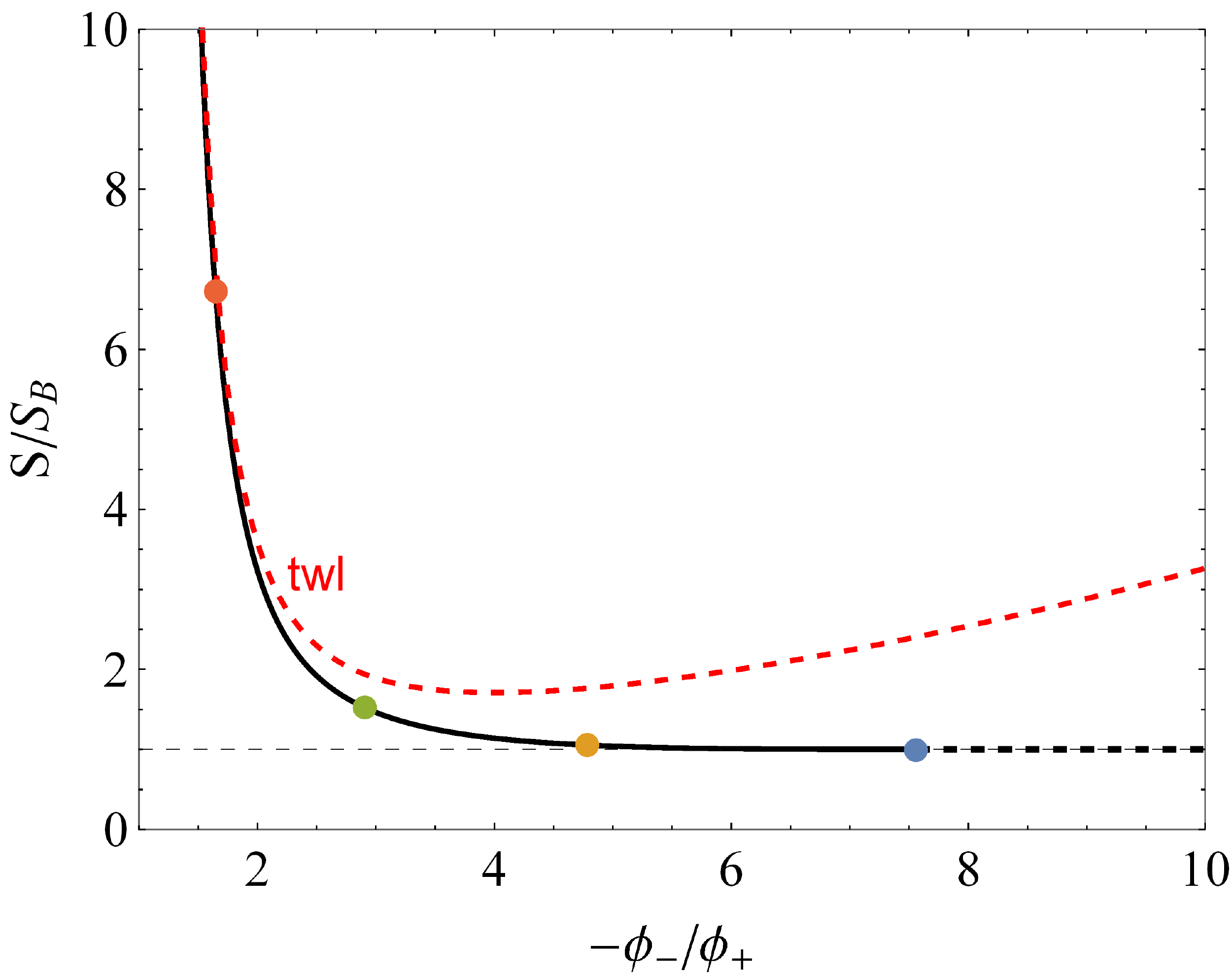}
\caption{For $V(\phi)=-m^2\phi^2/2$ with a false vacuum at $\phi_+=-1$. Upper left: Solid: Outer wall radius $r_+$ as
a function of the inner wall radius $r_-$  (both in units of $1/m$).
Dashed: approximation $m r_- \simeq m r_+ + \pi$.
Upper right: Profiles of the (pseudo)-bounces for different $\phi_-$. Black dots mark $r_\pm$ and the blue curve is the true bounce.
Lower left: Potential (black) and tunneling potentials for the previous (pseudo)-bounces (same color coding).
Lower right: Tunneling action for pseudo-bounces as $\phi_-$ is varied. Colored dots mark the previous (pseudo)-bounces. The red dashed line gives the improved thin-wall approximation of (\ref{m2tw}).
\label{fig:m2}
}
\end{figure}

If one considers $\phi_-<\phi_-^B$, there are pseudo-bounces
with profile 
\be
\Phi(r)=\left\{
\begin{matrix}
\phi_-\, , & r<r_- \\
{\displaystyle
\frac{\pi m\phi_-r_-^2}{2r}\left[J_{2-}Y_1(m r)-Y_{2-}J_1(m r)\right]}\, ,& r_- < r < r_+\\
\phi_+\, , & r>r_+
\end{matrix}
\right.
\label{psbounce}
\ee
where $Y_\alpha(x)$ is the Bessel function of the second kind and we use the short-hand notation
 $J_{2\pm}\equiv J_2(m r_\pm)$, $Y_{2\pm}\equiv Y_2(m r_\pm)$.
 Here, the inner and outer radii $r_\pm$  of a pseudo-bounce
satisfy the relations
\be
r_-^2\phi_-J_{2-} = r_+^2\phi_+J_{2+}\, ,\quad\quad
r_-^2\phi_-Y_{2-} = r_+^2\phi_+ Y_{2+}\ .
\label{rmrp}
\ee
which enforce
 [using $J_2(x)Y_1(x)-Y_2(x)J_1(x)=2/(\pi x)$]
\be
\Phi(r_\pm)=\phi_\pm\ , \quad \dot\Phi(r_\pm)=0\ .
\ee
It is understood that, for a given $r_-$, the outer radius $r_+(>r_-)$ is  the smallest possible solution of  $\dot\Phi(r_+)=0$, so that $\Phi(r)$ is monotonic in $(r_-,r_+)$. For $m r_-\gg 1$ (thin-wall pseudo-bounces) one has $m r_- \simeq m r_+ + \pi$.
The quantity $m r_+$ is plotted as a function of $m r_-$ in Fig.~\ref{fig:m2}, upper left plot. The same figure shows different pseudo-bounce profiles and the true bounce, in blue (upper right plot) and  the corresponding tunneling potentials, using the same color coding (lower left plot).

The Euclidean action in this case is
\bea
S_E&=&\left.\frac{\pi^2}{4}m^2r_+^4\phi_+^2\left\{1+\frac{\pi^2}{2}J_{2+}Y_{2+}\left[m^2r_+^2(J_{2+} Y_{2+}+J_{1+}Y_{1+})-2m r_+ (J_{1+}Y_{2+}+J_{2+}Y_{1+})\right]\right.\right.\nonumber\\
&-&\left.\frac{\pi^{3/2}}{2}  J_{2+}Y_{2+} G^{2,2}_{3,5}\left(m^2 r_+^2\left|\begin{matrix}
0,1/2,-1/2\\
0,2,-2,-1,-1/2
\end{matrix}
\right.\right)\right\}-\left\{
\begin{matrix}
r_+\rightarrow r_-\\
\phi_+\rightarrow \phi_-
\end{matrix}\right\}\ ,
\label{Sm2}
\eea
where $G$ is the Meijer function. It can be checked that 
this has the right limit for the proper bounce case (\ref{SB}).
Figure~\ref{fig:m2}, lower right plot, shows the tunneling action (\ref{Sm2}) as a function of $-\phi_-/\phi_+$, normalized to the bounce action (the colored dotted points correspond to the profiles shown in the same figure). The plot also shows the thin-wall approximation, with
\be
\sigma=\int_{\phi_+}^{\phi_-}\sqrt{2[V(\phi)-V_l(\phi)]}\simeq  \frac18 m\pi (\phi_--\phi_+)^2\ ,
\label{m2tw}
\ee
where we used the linear interpolation
$V_l(\phi)= m^2\left[\phi_+\phi_- -(\phi_++\phi_-)\phi\right]/2$
to define $\sigma$.
\begin{figure}[t!]
\begin{center}
\includegraphics[width=0.6\textwidth]{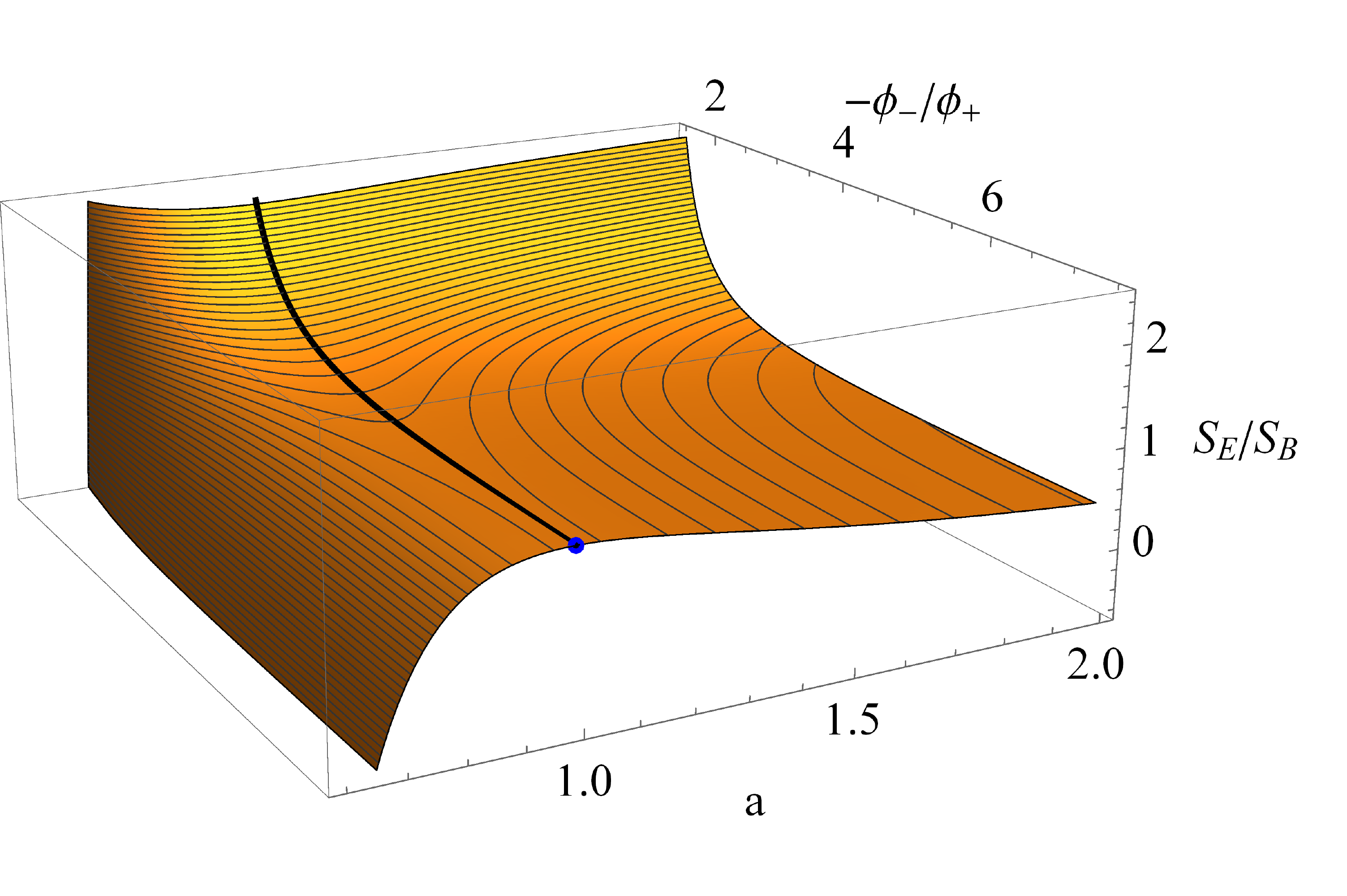}
\end{center}
\caption{Euclidean tunneling action for the potential $V(\phi)=-m^2\phi^2/2$ with $\phi_+=-1$ normalized to the bounce action in a slice of configuration space spanned by the pseudo-bounces $\Phi_{PB}(r)$ (parametrized by $\phi_-$) and an orthogonal direction parametrized by the rescaling parameter $a$, for configirations $\Phi_a(r)=\Phi_{PB}(a r)$. The black line follows the trajectory of pseudo-bounces, which ends at the bounce, marked by the blue dot.
\label{fig:ridgem2}
}
\end{figure}

We can compare the behaviour of the action with the one for the negative-quartic potential of section~\ref{sec:lambdaphi4}, shown in Fig.~\ref{fig:SE}. We see that now the action reaches its minimum at a finite $\phi_-$, corresponding to the bounce solution
(blue dot in the figure). If we set the minimum at $\phi_->\phi_-^B$ tunneling will nevertheless proceed via the proper bounce towards $\phi_-^B$. From the discussion in section~\ref{sec:wkb} it should also be clear that tunneling configurations with $\phi_->\phi_-^B$
can be realized, but their action cost is higher than the bounce minimal one.

We will use this example to illustrate one further point. The Euclidean action has a negative mode so that the bounce is not a minimum of $S_E$ but a saddle point and the same negative mode appears for pseudo-bounces. We can see this most easily by considering
rescaled configurations $\phi_a(r)\equiv \phi_{PB}(a r)$, where
$\phi_{PB}(r)$ is a pseudo-bounce (or a proper bounce), that keep $\phi_{PB}(0)$ fixed. By rescaling the Euclidean coordinates and using Derrick's theorem we get the scaling
\be
S_E[\phi_a]=\frac{1}{a^2}\left(2-\frac{1}{a^2}\right)S_E[\phi_{PB}]\ .
\ee
We plot in Fig.~\ref{fig:ridgem2} the same action as in Fig.~\ref{fig:m2} but opening up configuration space along the rescaling parameter $a$. This shows that the pseudo-bounce trajectory can be considered as the bottom of a valley for the Euclidean action only if one removes the negative mode. Including that mode one sees that the trajectory follows the top of a ridge. This is the root of some of the difficulties with contsrained instantons or new valley methods
we discussed at the end of last section. One significant advantage
of the tunneling potential approach is that the action has no such negative mode.

\section{Potential $\bma{V(\phi)=\frac{1}{2}m^2\phi^2-\frac14\lambda\phi^4}$\label{sec:m2lambdaphi4}}

Finally, we consider a potential that combines those of previous sections
\be
V(\phi)=\frac12 m^2\phi^2-\frac14\lambda\phi^4\ .
\ee
We fix the false vacuum at $\phi_+=0$, study tunneling towards $\phi_->0$ and consider $m^2$ (of either sign)
as the sole parameter breaking scale invariance. By the scaling argument used in the Introduction we know this potential does not have a bounce solution describing the decay of the $\phi_+$ false vacuum. In particular, for $m^2>0$ all trial solutions of the Euler-Lagrange equation~(\ref{EoM4}) are undershots, while for $m^2<0$
all are overshots.

By the same reasoning we used in section~\ref{sec:lambdaphi4}, we deduce that the tunneling action must be a function of the ratio $\phi_-^2/m^2$:
\be
S=S(\phi_-^2/m^2)\ ,
\ee
with
\be
S(\infty)=\frac{8\pi^2}{3\lambda}\ ,\quad S(2/\lambda)=\infty\ .
\ee
The last case corresponds to the thin-wall limit with $\phi_-^2=2m^2/\lambda$ (with $m^2>0$) that gives degenerate vacua at $\phi_\pm$. We expect, at least for $m^2>0$, a behaviour similar to the one found in Sect.~\ref{sec:lambdaphi4}: pseudo-bounce solutions with non-zero $r_-$
and a tunneling action monotonically decreasing towards the no-scale value $8\pi^2/(3\lambda)$ when $\phi_-^2/m^2\rightarrow \infty$. An analytical solution does not seem feasible, although one could try a perturbative approach treating $m^2$ as a small perturbation of the no-scale case, in analogy to the analyses in \cite{NewValley,NN,Matthew}, that used constrained instantons or the new valley method. Alternatively one could simply use, at least for $m^2>0$, numerical solutions to solve the Euler-Lagrange equation (\ref{EoM4}), implementing the need of nonzero $r_-$ to find the pseudo-bounces.

Whatever we do, the results for $m^2>0$ are qualitatively similar to the ones obtained in section~\ref{sec:lambdaphi4}, so we will instead resort to a much simpler analysis based on the Minkowskian approach, as this will also serve to clarify the case with $m^2<0$. 
Following the discussion in section~\ref{sec:wkb}, we consider two paths in configuration space that connect the $\phi_+$ false vacuum
and a configuration that contains a zero-energy tunneling bubble.
(For a related discussion see Sect.~7 of \cite{Matthew}).

The first path has a Fubini profile
\be
\Phi_\alpha(r)=\frac{\phi_-}{1+(r^2+\alpha^2)/R^2}\ ,\quad 
\frac{1}{R^2}=\frac{\lambda\phi_-^2}{8}-2m^2 \ ,
\label{FubiniPath}
\ee
while for the second path we take a Gaussian profile:
\be
\Phi_\alpha(r)=\phi_-\ e^{-(r^2+\alpha^2)/R^2}\ ,\quad 
\frac{1}{R^2}=\frac{\lambda\phi_-^2}{12\sqrt{2}}+\frac{m^2}{3} \ .
\label{GaussianPath}
\ee
As usual, $\alpha=\infty$ corresponds in both cases to the $\phi_+=0$ vacuum, while for $\alpha=0$ we have the tunneling bubble configuration. The radius $R$ has been chosen to ensure the bubble has zero energy. When the sign of $m^2$ can lead to $R^2<0$ we assume $|m^2|$ is small enough so that $R$ is real.

For these two paths we can then calculate the WKB tunneling action
as explained in Sect.~\ref{sec:wkb} in the understanding that the action will be even lower for the true tunneling path. For the Gaussian path one can perform the WKB integrals numerically.
For the Fubini path we can calculate the WKB action analytically, getting
\be
S_\pm=\frac{8\pi^2}{3\lambda}f_\pm\left(\frac{\lambda\phi_-^2}{16|m^2|}\right)\ , \quad
\ee
where the subindex $+ (-)$ corresponds to $m^2>0$ ($m^2<0$). 

\begin{figure}[t!]
\begin{center}
\includegraphics[width=0.6\textwidth]{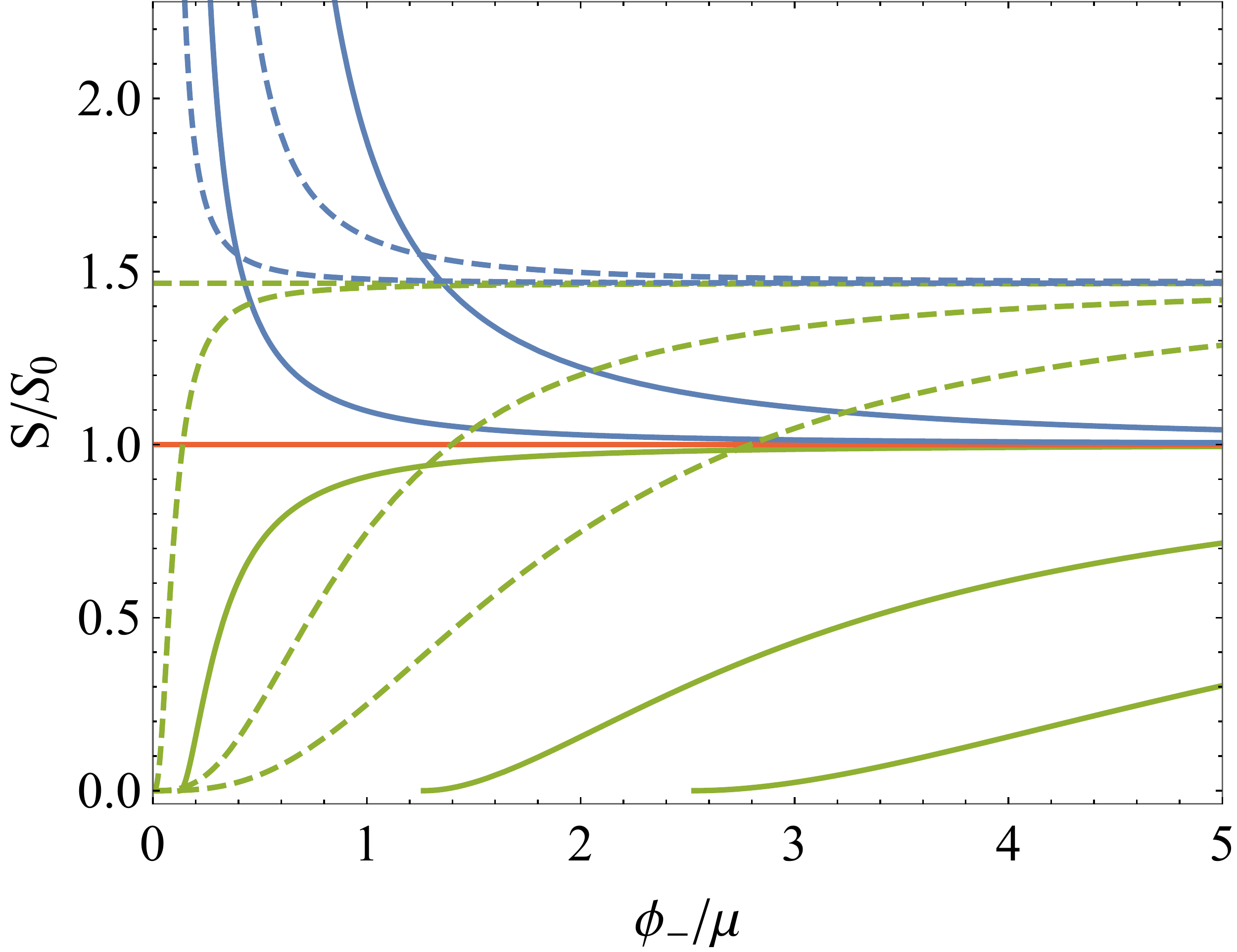}
\end{center}
\caption{ WKB tunneling actions for 
for the potential $V(\phi)=-\lambda\phi^4/4+m^2\phi^2/2$, with $\lambda=0.1$ and different masses,
along two different tunneling paths
in configuration space as a function of the tunneling end-point $\phi_-$. Dashed lines: Gaussian paths as in (\ref{GaussianPath}).
Solid lines: Fubini paths as in (\ref{FubiniPath}). We take 
 $m^2/\mu^2=\{10^{-3},10^{-4},0,-10^{-4},-10^{-2},-4\times 10^{-2}\}$ where $\mu$ is some arbitrary reference scale. The curves deviate from the flat $m^2=0$ ones in proportion to the size of $m^2$, upwards for $m^2>0$ and downwards for $m^2<0$. \label{fig:m2impact}
}
\end{figure}

For $m^2>0$, 
one gets [with $x=\lambda\phi_-^2/(16m^2)> 1$]
\be
f_+(x)=\frac{\left[
(x-1)E(-x)+(x+1)K(-x)\right]}{(x-1)^{3/2}}\ ,
\ee
where $K$ and $E$ are the complete elliptic integrals of the first and second kinds, respectively.
This function has the asymptotic values $f_+(\infty)=1$
(when $m^2$ is not relevant and one recovers the Fubini action) and
$f_+(1)=\infty$ [for this type of profile the lower limit $\phi_->4m/\sqrt{\lambda}$ for tunneling to be allowed is stronger than for the thin wall case ($\phi_->m\sqrt{2/\lambda}$)].

For $m^2<0$, instead, one gets
\be
f_-(x)=\sqrt{x}\frac{\left[
(x+1)E(1-1/x)-2K(1-1/x)\right]}{(x+1)^{3/2}}\ ,
\ee
with $x=-\lambda\phi_-^2/(16m^2)> 1$. Now $f_-(1)=0$ and $f_-(\infty)=1$.

 We show the results for both types of path in Fig.~\ref{fig:m2impact}. Solid (dashed) lines correspond to the 
Fubini (Gaussian) path, for a number of values of $m^2$ as indicated.   Consider first the case $m^2>0$ (curves with negative slope). We see that Gaussian profiles asymptote for $\phi_-\rightarrow\infty$ to a value higher than
the Fubini ones, for which one recovers the expected no-scale result
$S_0=8\pi^2/(3\lambda)$. For the dependence of the tunneling action on $m^2$, we find that higher values of $m^2$ increase the action as the potential barrier grows. The action for the pseudo-bounce path will lie below the curves shown but with a similar behaviour, asymptoting to $S_0$ too.

The case $m^2<0$ (curves with positive slope) is qualitatively different. For both types of path we see that the action can be made arbitrarily small for small enough $\phi_-$. This means that the tunneling rate is not suppressed at all and the vacuum is completely unstable.

\section{Examples of analytical $\bma{V_t}$\label{sec:analyticVt}}

In the tunneling potential approach, for a given potential $V(\phi)$
one should find the corresponding tunneling potential $V_t(\phi)$. In \cite{E} it was shown how to solve the inverse problem of finding $V$ corresponding to a postulated $V_t$, which is easier to do as the differential equation (\ref{EoMVt}) is linear and first order in $V$. Solving this inverse problem was useful to construct special potentials for which the tunneling problem could be solved entirely analytically.

While that was done for the tunneling potentials corresponding to true Euclidean bounces, in this section we consider this inverse problem for ``restricted'' tunneling potentials, those corresponding
to Euclidean pseudo-bounces. The general formula for $V(\phi)$
in terms of $V_t(\phi)$  used in \cite{E} involves inverse powers of $V_t'(\phi_0)$. As this derivative vanishes for pseudo-bounces, the formula needs to be modified. It is straightforward to 
do so and get instead 
\be
V(\phi)=[V_t'(\phi)]^2\left\{\frac{V_c}{[V_t'(\phi_c)]^2}+\int_{\phi_c}^{\phi}\frac{4[V_t'(\bar\phi)]^2-6V_t(\bar\phi)V_t''(\bar\phi)}{3[V_t'(\bar\phi)]^3}d\bar\phi
\right\}\ ,\label{Van}
\ee
where $\phi_c\neq \phi_-$ is some field value in the interval $(\phi_+,\phi_-)$
and $V_c$ is some arbitrary constant that fixes the value of $V(\phi_c)=V_c$. The formula in \cite{E} is recovered for $\phi_c=\phi_0$ and $V_c=V(\phi_0)$.

In \cite{E} the strategy was to postulate a simple (monotonically decreasing) $V_t(\phi)$ hoping to get a simple enough $V(\phi)$. For restricted $V_t$'s we rather postulate a simple form for $V_t'$ as we know this has to be a negative function that vanishes at $\phi_\pm$. The point $\phi_-$ is now a regular singular point of the differential equation (\ref{EoMVt}), as
\be
V_t'' = \frac{(V'-4V_t'/3)V_t'}{2(V-V_t)}\ ,
\ee
and $(V'-4V_t'/3)$ does not vanish at $\phi_-$ for pseudo-bounces.
Therefore, the tunneling potential $V_t(\phi)$
cannot be expanded in a Taylor series around $\phi_-$ as $V_t''$ diverges there.

\begin{figure}[t!]
\includegraphics[width=0.5\textwidth]{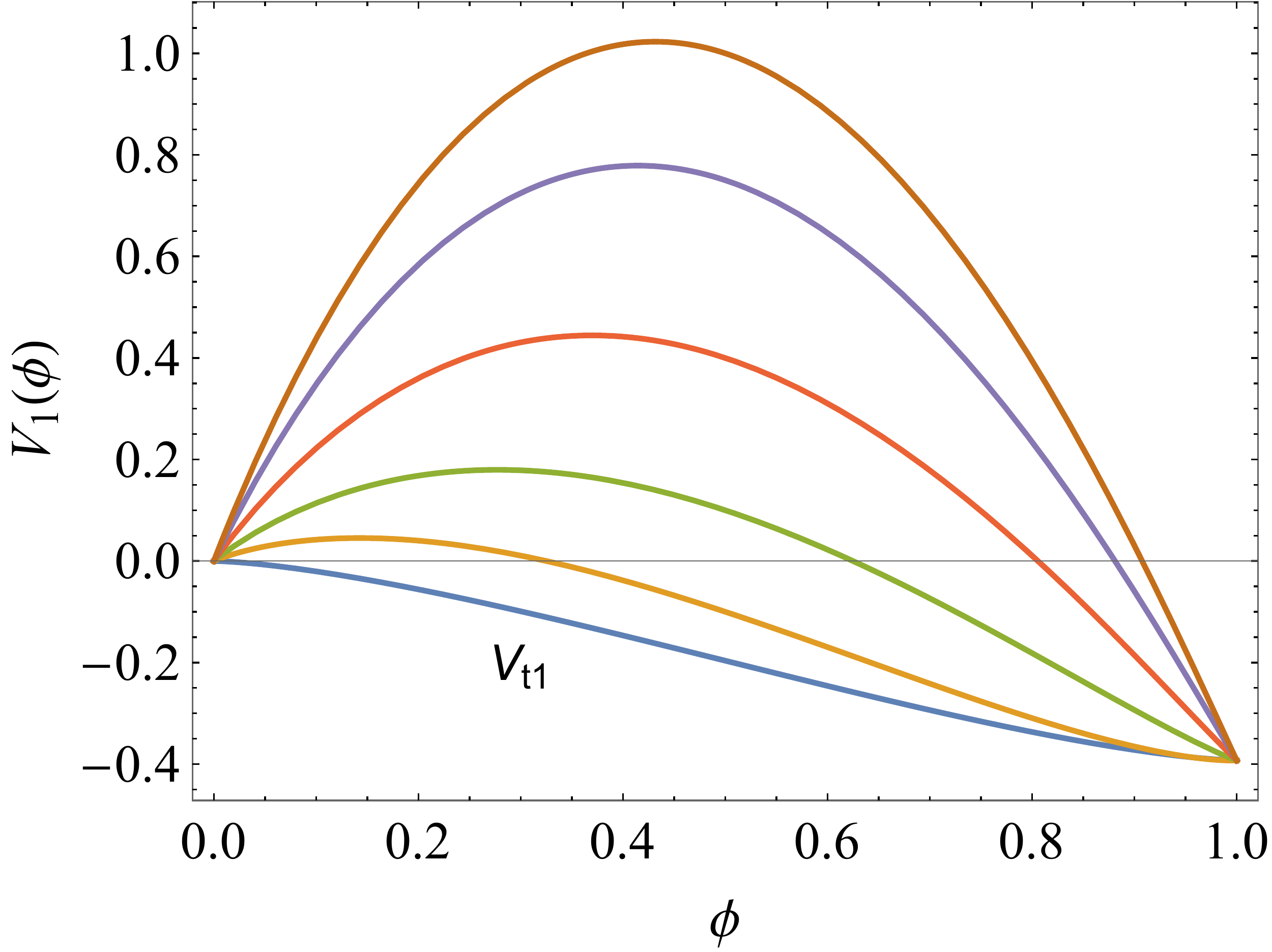}\,\,\,
\includegraphics[width=0.5\textwidth]{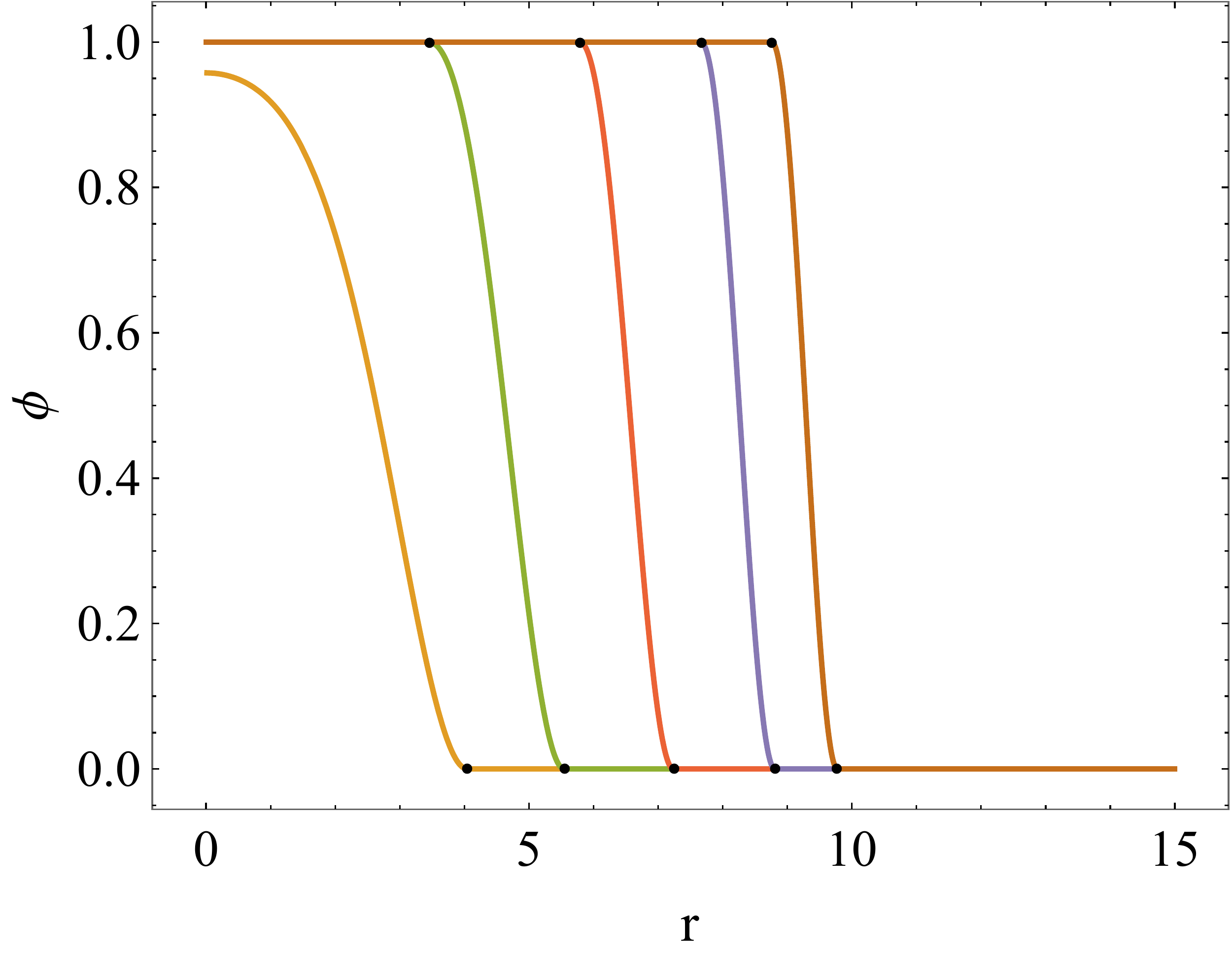}
\caption{Left: Different potentials $V_1(\phi)$ of (\ref{V1}) for the tunneling potential $V_{t1}(\phi)$ of (\ref{Vt1}) with different values
of $V(1/2)$. Right: Pseudo-bounce profiles corresponding to the potentials on the left, with the same color coding. The radii $r_\pm$
are marked by black dots.
\label{fig:Van1}
}
\end{figure}

Taking the previous property into account we consider as our first example the simple choice
\be
V_{t1}'(\phi)=-\sqrt{\phi(1-\phi)}\ ,
\ee
defined for the interval $(\phi_+,\phi_-)=(0,1)$ (dimensionful constants can be introduced easily if needed). This leads by direct integration to
\be
V_{t1}(\phi)= \frac14 \left[(1-2\phi)\sqrt{\phi(1-\phi)}-\arcsin\sqrt{\phi}\right]\ ,
\label{Vt1}
\ee
with an arbitrary integration constant chosen so that $V_{t1}(0)=0$.
From the formula (\ref{Van}) one then gets
\be
V_1(\phi)=\phi(1-\phi)\left[\left.\frac{}{}F_1(x)\right|_{\phi_c}^\phi+\frac{V_c}{\phi_c(1-\phi_c)}\right]\ ,
\label{V1}
\ee
where
\be
F_1(x)=\frac{1}{4x(1-x)}\left\{(1-2x)\sqrt{x(1-x)}+\left[1+\frac{8}{3}x(1-x)\right]\arccos\sqrt{x}-\frac{\pi}{2}\right\}\ .
\ee
Fig.~\ref{fig:Van1}, left plot, shows different potentials $V_1(\phi)$ for different choices of $V_c=V(1/2)$. The profile $\phi(r)$ of the corresponding pseudo-bounces can be recovered from $V_1$ and $V_{t1}$ [as the inverse function $r(\phi)$ is determined by Eq.~(\ref{r})] and is shown, using the same color coding, in the right plot. The black dots mark the radii $r_\pm$. The lowest potential (and profile) correspond in fact to a proper bounce: the intersection between $V$ and $V_t$ takes place at $\phi_-<1$, where $V_t'\neq 0$. So we see that the analytical formula can interpolate smoothly between pseudo-bounce and bounce solutions. Notice that $r_+$ is finite for all cases; this is due to the potential having a non-zero derivative arbitrarily close to the false minimum at the origin. This can be easily modified, as we do in the next example, by choosing a $V_t'$ that can be expanded in a Taylor series around $\phi_+=0$.

Consider then 
\be
V_{t2}'(\phi)=-\phi\sqrt{1-\phi}\ ,
\ee
which leads by integration to
\be
V_{t2}(\phi)= \frac{2}{15} \left[(2+3\phi)(1-\phi)^{3/2}-2\right]\ .
\label{Vt2}
\ee
From the formula (\ref{Van}) one then gets
\be
V_2(\phi)=\phi^2(1-\phi)\left[\left.\frac{}{}F_2(x)\right|_{\phi_c}^\phi+\frac{V_c}{\phi_c^2(1-\phi_c)}\right]\ ,
\label{V2}
\ee
where
\be
F_2(x)=\frac{2}{3}\left\{\frac{1}{5x^2}\left[
(2+3x)\sqrt{1-x}-\frac{2}{1-x}\right]+\mathrm{arctanh}\sqrt{1-x}\right\}\ .
\ee

\begin{figure}[t!]
\includegraphics[width=0.5\textwidth]{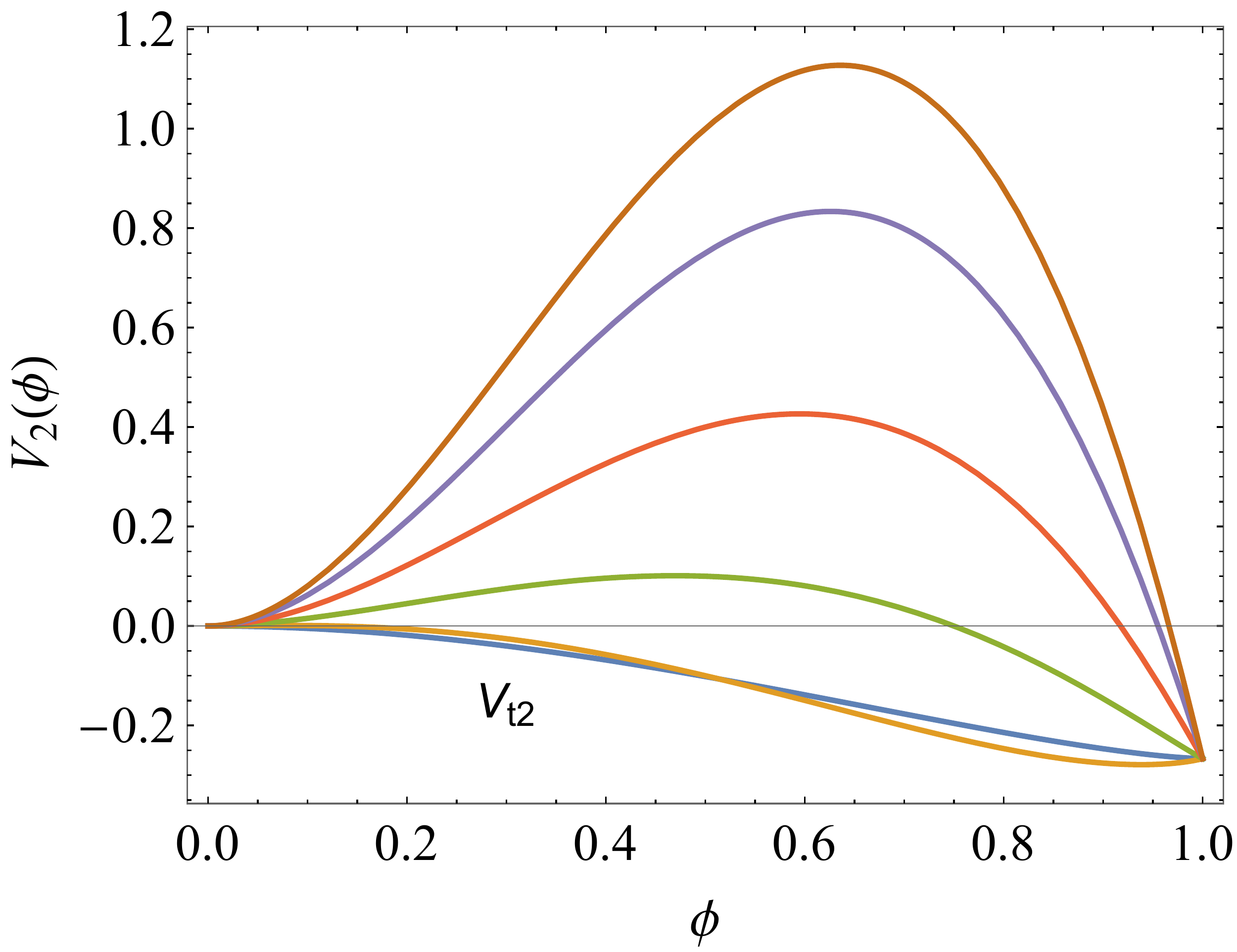}\,\,\,
\includegraphics[width=0.5\textwidth]{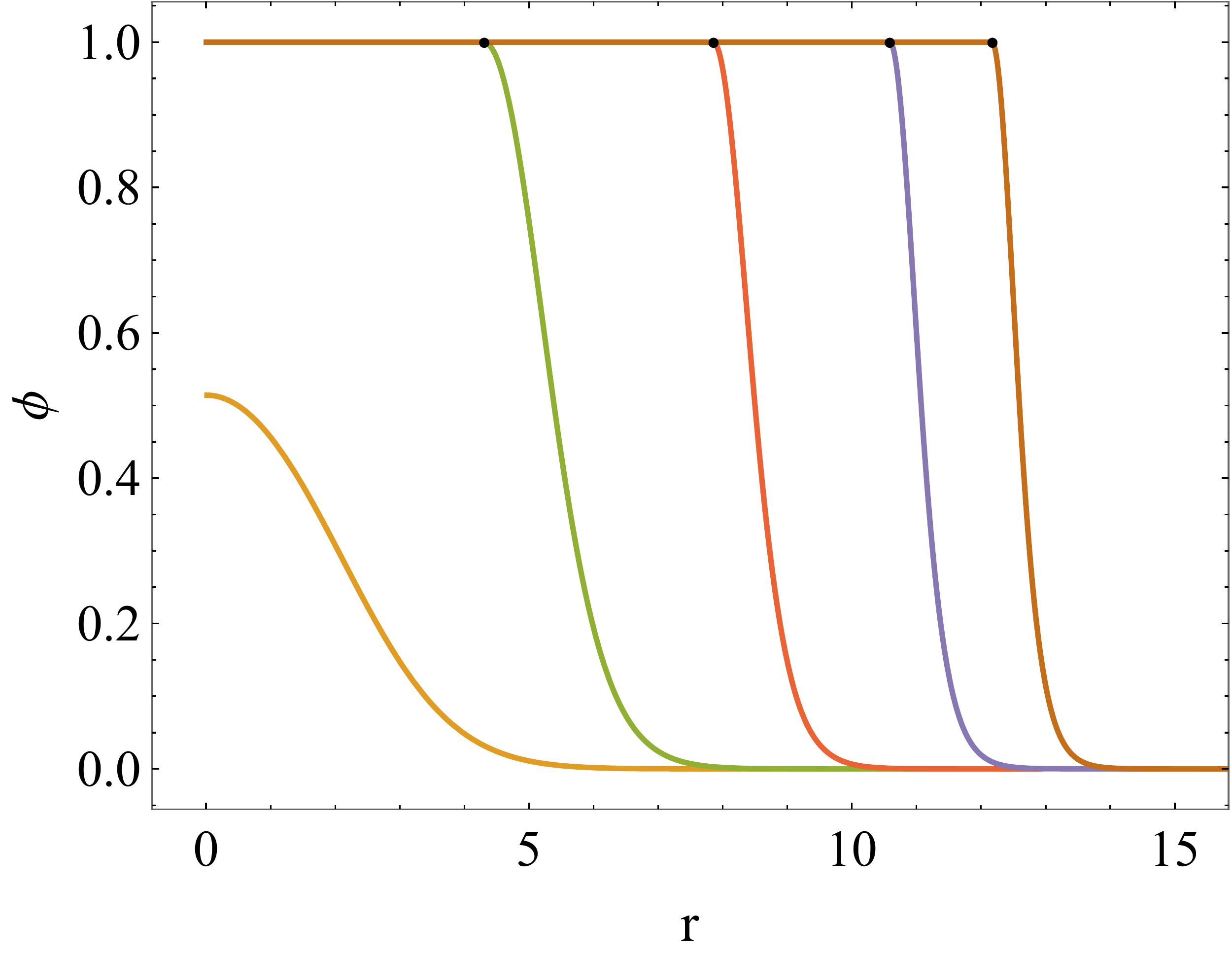}
\caption{Left: Different potentials $V_2(\phi)$ of (\ref{V2}) for the tunneling potential $V_{t2}(\phi)$ of (\ref{Vt2}) with different values
of $V(1/2)$. Right: Pseudo-bounce profiles corresponding to the potentials on the left, with the same color coding. The radii $r_\pm$
are marked by black dots.
\label{fig:Van2}
}
\end{figure}

The $V_2(\phi)$ potentials and corresponding (pseudo-)bounces are shown in Fig.~\ref{fig:Van2}, for different choices of $V_c=V(1/2)$. The black dots mark the corresponding $r_-$ while $r_+\rightarrow \infty$ (the false vacuum does have zero derivative now). As in the previous example, the lowest potential (and profile) correspond to a proper bounce with $\phi_-<1$.

\section{Summary and outlook\label{sec:conclusions}}

By using a combination of methods we have shown that false vacua that cannot decay via proper bounces would still decay and described in detail how they do it. 
In the Euclidean approach the decay is mediated by Euclidean configurations we call pseudo-bounces, which have a homogeneous core where the field sits at a given field value $\phi_-$, that probes the deeper parts of the potential, up to a radial 
distance $r_-$. Beyond that inner radius the field transitions smoothly to the false vacuum $\phi_+$. In that wall region the Euclidean Euler-Lagrange equation (\ref{EoM4}) is satisfied (while it is not in the core). Nevertheless the pseudo-bounce inherits some of the good properties of proper bounces (except extremizing the Euclidean action). In particular: {\em 1)} the pseudo-bounce Euclidean action can be expressed as half the action coming from the field gradient in the wall region, and {\em 2)} the slice of the pseudo-bounce 
at zero Euclidean time is a three-dimensional bubble configuration of zero energy. We gave concrete examples of these pseudo-bounce configurations in the text but the properties above hold in general
when there is no bounce.

The Euclidean analysis is ultimately justified by the Minkowski approach. We construct paths in configuration space that join the false vacuum with a zero-energy tunneling bubble configuration
(corresponding to the Euclidean zero-time slices). For those paths we then calculate the minimal WKB tunneling exponent, finding agreement with the previous Euclidean result.

Finally, the most natural approach uses the tunneling potential method of \cite{E}. For a fixed value of $\phi_0$ in the deeper regions of the potential, one finds the tunneling potential $V_t(\phi)$ that minimizes the simple new action (\ref{SVt}). Such  ``restricted'' $V_t$ gives the minimal action for fixed $\phi_0$ (while the true $V_t$ would minimize the action when $\phi_0$ is free to vary). Such restricted $V_t$ is also a solution of the corresponding Euler-Lagrange equation (\ref{EoMVt}).
The only difference with respect to a proper $V_t$ is the boundary condition $V_t'(\phi_0)=0$ [instead of $V_t'(\phi_0)=3V'(\phi_0)/4$]. 
The action obtained in this way agrees with the previous ones and the field profile derived from $V_t$ reproduces the pseudo-bounce profile. This method also allows to find easily a simple expression for the variation of the action when $\phi_0$ is varied, as given by (\ref{dSdphi}).

Whichever method is used, the picture one  gets for the decay of false vacua without bouce is of a nearly flat direction of the action functional in field configuration space. That flat direction (or valley bottom) is parametrized naturally and faithfully by the value of $\phi_0$ and consists of a family of pseudo-bounce configurations with the lowest values of the tunneling action (at fixed $\phi_0$). 

To finally calculate the decay rate one should integrate along the valley bottom using the collective coordinate method (See e.g. \cite{OldValley2,NewValley,Matthew}). In particular cases of physical interest, renormalization effects or UV physics can play a very important role in changing the shape of the valley and therefore in getting the final rate. In this paper we have focused on the pre-requisite need of tracing accurately that valley bottom in general. Model dependent effects can be added on top, case by case.

\section*{Acknowledgments\label{sec:ack}} 
I thank Pepe Barb\'on for interesting discussions.
This work has been supported by the ERC
grant 669668 -- NEO-NAT -- ERC-AdG-2014, the Spanish Ministry MINECO under grants  2016-78022-P and
FPA2014-55613-P and the grant SEV-2016-0597 of the Severo Ochoa excellence program of MINECO .


\begin{thebibliography}{99}

\bibitem{SM}
  N.~Cabibbo, L.~Maiani, G.~Parisi, R.~Petronzio,
  Nucl. Phys. B {\bf 158} (1979) 295;
  N.~Krasnikov,
  Yad. Fiz.  {\bf 28} (1978) 549;
  P.~Hung,
  Phys. Rev. Lett.  {\bf 42} (1979) 873;
  G.~Degrassi, S.~Di Vita, J.~Elias-Mir\'o, J.~Espinosa, G.~Giudice, G.~Isidori, A.~Strumia,
  JHEP {\bf 1208} (2012) 098
  \arXiv{1205.6497}{ph};
  D.~Buttazzo, G.~Degrassi, P.~Giardino, G.~Giudice, F.~Sala, A.~Salvio, A.~Strumia,
  JHEP {\bf 1312} (2013) 089
  \arXiv{1307.3536}{ph};
  A.~Bednyakov, B.~Kniehl, A.~Pikelner, O.~Veretin,
  Phys. Rev. Lett. {\bf 115} (2015) 201802
  \arXiv{1507.08833}{ph}.

\bibitem{CMS}
  S.~Chigusa, T.~Moroi and Y.~Shoji,
  Phys.\ Rev.\ Lett.\  {\bf 119} (2017) 211801
  \arXiv{1707.09301}{ph};
  Phys.\ Rev.\ D {\bf 97} (2018) 116012
  \arXiv{1803.03902}{ph}.
 

\bibitem{Matthew}
  A.~Andreassen, W.~Frost and M.~Schwartz,
  Phys.\ Rev.\ D {\bf 97} (2018) 056006
  \arXiv{1707.08124}{ph}.
     
\bibitem{Coleman}
  S.~Coleman,
  Phys.~Rev.~D {\bf 15} (1977) 2929
   Erratum: [Phys. Rev. D {\bf 16} (1977) 1248].


\bibitem{CGM}
  S.~Coleman, V.~Glaser, A.~Martin,
  Commun. Math. Phys.  {\bf 58} (1978) 211.
  
\bibitem{Fubini}
  S.~Fubini,
  Nuovo Cim.\ A {\bf 34} (1976) 521.

\bibitem{Lipatov}
  L.~Lipatov,
  Sov.\ Phys.\ JETP {\bf 45} (1977) 216
   [Zh.\ Eksp.\ Teor.\ Fiz.\  {\bf 72} (1977) 411].


\bibitem{Arnold}
  P.~Arnold,
  Phys.\ Rev.\ D {\bf 40} (1989) 613;
  P. Arnold and S. Vokos,
  Phys.\ Rev.\ D {\bf 44} (1991) 3620.

  
\bibitem{NoBounce}
  J.~Gonz\'alez, A.~Bellor\'{\i}n, M.~Garc\'{\i}a-\~{N}ustes, L.~Guerrero, S.~Jim\'enez, J.~Mar\'{\i}n and L.~V\'azquez,
  JCAP {\bf 1806} (2018)  033
  \arXiv{1710.05334}{th}.
  
\bibitem{Affleck}
  I.~Affleck,
  Nucl.\ Phys.\ B {\bf 191} (1981) 429.
  
\bibitem{FY}
  Y.~Frishman and S.~Yankielowicz,
  Phys.\ Rev.\ D {\bf 19} (1979) 540.
  
\bibitem{OldValley1}
  E.~Shuryak,
  Nucl.\ Phys.\ B {\bf 302} (1988) 621.
\bibitem{OldValley2}
  I.~Balitsky and A.~Yung,
  Phys.\ Lett.\ B {\bf 168} (1986) 113;
  A.~Yung,
  Nucl.\ Phys.\ B {\bf 297} (1988) 47.
\bibitem{OldValley3}
  V.~Khoze and A.~Ringwald,
  Phys.\ Lett.\ B {\bf 259} (1991) 106.
 
    
\bibitem{NewValley}
  H.~Aoyama and H.~Kikuchi,
  Nucl.\ Phys.\ B {\bf 369} (1992) 219;
  H.~Aoyama and S.~Wada,
  Phys.\ Lett.\ B {\bf 349} (1995) 279
\arXivold{th/9408156};
  H.~Aoyama, T.~Harano, M.~Sato and S.~Wada,
  Nucl.\ Phys.\ B {\bf 466} (1996) 127
  \arXivold{th/9512064};
  H.~Aoyama, T.~Harano, H.~Kikuchi, M.~Sato and S.~Wada,
  Phys.\ Rev.\ Lett.\  {\bf 79} (1997) 4052
  \arXivold{th/9606159}.
  

\bibitem{E}
  J.~Espinosa,
~JCAP\,{\bf 07}~(2018)~36,~\arXiv{1805.03680}{th}.
   

   
\bibitem{BBW}
  T.~Banks, C.~Bender, T.~Wu,
  Phys.\ Rev.\ D {\bf 8} (1973) 3346.
  
\bibitem{BC}
  K.~Bitar and S.~Chang,
  Phys.\ Rev.\ D {\bf 17} (1978) 486;
  Phys.\ Rev.\ D {\bf 18} (1978) 435.

\bibitem{Gino}
  L.~Di Luzio, G.~Isidori and G.~Ridolfi,
  Phys.\ Lett.\ B {\bf 753} (2016) 150
  \arXiv{1509.05028}{hep-ph}.
  
    
\bibitem{Eg}
  J.~Espinosa,
  \arXiv{1808.00420}{th}.
 
\bibitem{EK}
  J.~Espinosa and T.~Konstandin,
  JCAP {\bf 1901} (2019) 051
  \arXiv{1811.09185}{th}.

\bibitem{Derrick}
  G.~Derrick,
  J. Math. Phys.  {\bf 5} (1964) 1252.
    
\bibitem{NN}
  M.~Nielsen and N.K.~Nielsen,
  Phys.\ Rev.\ D {\bf 61} (2000) 105020
  \arXivold{th/9912006}.
   

\end{thebibliography}
\end{document}